\documentclass[11pt,reqno]{amsart}
\usepackage{multirow}
\usepackage[centertags]{amsmath}
\usepackage{amsfonts}
\usepackage{amssymb}
\usepackage{amsthm}
\usepackage{newlfont}
\usepackage{a4}
\usepackage{fullpage}
\usepackage{enumerate}
\usepackage[pdftex]{graphicx,color}
\theoremstyle{definition}

\theoremstyle{remark}
\newtheorem{example}{Example}[section]
\usepackage{bbm}

\newcommand{\id}{\mathfrak{1}}

\newcommand{\map}{\rightarrow}

\newcommand{\q}{\quad}

\renewcommand{\epsilon}{\varepsilon}

\newcommand{\ep}{\varepsilon}
\newcommand{\la}{\lambda}
\newcommand{\al}{\alpha}
\newcommand{\om}{\omega}
\renewcommand{\rho}{\varrho}
\renewcommand{\phi}{\varphi}

\newcommand{\R}{{\mathbb{R}}}

\newcommand{\N}{{\mathbb N}}

\newcommand{\Z}{\mathbb{Z}}
\newcommand{\C}{\circ}

\newcommand{\set}[2]{\left\{#1 \, |\, #2 \right\}}
\newcommand{\setb}[2]{\left\{#1 \, \mid\, #2 \right\}}

\newcommand{\abs}[1]{\left\vert#1\right\vert}
\newcommand{\wt}{\widetilde}

\newcommand{\sca}[2]{\langle #1,\, #2\rangle}

\def\R{\mathbb R}
\def\Z{\mathbb Z}
\def\N{\mathbb N}
\def\C{\mathbb C}

\def\C{\mathbb C}
\def\R{\mathbb R}
\def\Z{\mathbb Z}
\def\N{\mathbb N}

\def\wt{\widetilde}

\def\o{\omega}

\begin{document}

\title
{Six types of $E-$functions of the Lie groups $O(5)$ and $G(2)$}

\author{Lenka H\'akov\'a$^{1}$}
\author{Ji\v{r}\'{\i} Hrivn\'{a}k$^{1,2}$}
\author{Ji\v{r}\'{\i} Patera$^{1,3}$}

\begin{abstract}\ 

New families of $E$-functions are described in the context of the compact simple Lie groups $O(5)$ and $G(2)$. These functions of two real variables generalize the common exponential functions and for each group, only one family is currently found in the literature. All the families are fully characterized, their most important properties are described, namely their continuous and discrete orthogonalities and decompositions of their products.

\end{abstract}
\maketitle

\noindent
$^1$ Centre de Recherches Math\'ematiques et D\'epartement de Math\'ematiques et de Statistique,
         Universit\'e de Montr\'eal,
         C.~P.~6128 -- Centre Ville,
         Montr\'eal, H3C\,3J7, Qu\'ebec, Canada;\\ patera@crm.umontreal.ca, hakova@dms.umontreal.ca\\
$^2$ Department of Physics, Faculty of Nuclear Sciences and
Physical Engineering, Czech Technical University in Prague,
B\v{r}ehov\'a~7, 115 19 Prague 1, Czech Republic;
jiri.hrivnak@fjfi.cvut.cz\\
$^3$ MIND Research Institute,
         3631 S. Harbor Blvd., Suite 200,
Santa Ana, CA 92704,  USA
\date{\today}

\maketitle
\section{Introduction}
We consider six infinite families of special functions which can be derived from the compact simple Lie groups of types
\begin{gather}\label{thegroups}
B_n\,,\quad C_n\,,\quad G_2\,,\quad F_4\,,\qquad 2\leq n<\infty\,.
\end{gather}
These are all the simple Lie groups with roots of two different lengths. One of the six families has been described previously \cite{KP1}.  The other families are new. 

Within each of the six family,  the functions
 \newline
(i) depend of $n$ real variables, $n$ being the rank of the underlying simple Lie group;
 \newline
(ii) are periodic in various ways over the entire real Euclidean space $\R^n$;
 \newline
(iii) are pairwise orthogonal when integrated over a finite (`fundamental') region of $\R^n$;
 \newline
(iv) are pairwise orthogonal when their values, sampled  at lattice fragment in the fundamental region, are summed up, the lattice being of any density and of symmetry dictated by the underlying group \eqref{thegroups}.

We limit our considerations to the Lie groups \eqref{thegroups} of rank two, namely the groups $O(5)\equiv Sp(4)$ and $G(2)$. Their Lie algebras are denoted $B_2\equiv C_2$ and $G_2$. There are neither principal nor technical obstacles to generalize the results of this paper to simple Lie groups of type \eqref{thegroups} of ranks greater than 2.

A basic tool for defining the functions are the particular subgroups $W^e, W^s, W^l$ of the finite reflection group $W$, which is the symmetry group of the root systems of the groups \eqref{thegroups}. These subgroups are called the even subgroups of $W$. They are of index 2 in $W$ and it is seen from the orders of the reflection groups that they are not reflection generated groups. 

In addition to the well known and extensively studied \cite{KP2,KP3} symmetric ($C$-~) and antisymmetric ($S$-~) functions of the Weyl group orbit, new families of $W$-orbit functions for the groups \eqref{thegroups}, called $S^L$-~ and $S^S$-~, were recently discovered. In \cite{P}, functions defined using the subgroup $W^e$, called $E$-functions, were studied. Those functions are defined for every compact simple Lie group \cite{KP1,PK} and they can be written as sums of symmetric and antisymmetric orbit functions (symbolically written as $E=C+S$). By considering the sums of pairs of the functions $C$, $S$, $S^L$, and $S^S$ built using the same Weyl group we obtain 6 families of $E-$functions:  
 We use the following notation for those functions 
\begin{alignat}{2}\label{ecka}
\Xi^{e+} &= C+S,  \qquad \qquad&&  \Xi^{e-}= S^L+S^S,\notag\\
\Xi^{s+} &= C+S^S,\qquad \qquad&&  \Xi^{s-}= S+S^S,\\
\Xi^{l+} &= C+S^L,\qquad \qquad&&  \Xi^{l-}= S+S^L\,. \notag
\end{alignat}
In this paper we use a different approach to construct the $E-$functions using the sign homomorphisms (section~\ref{homo}). 
 
General motivation \cite{R} for the study of $E$-functions of two variables resides in the processing of digitally given data. The advantage of having a larger choice of families of functions is the fact that they are orthogonal in regions of different shapes, which can be more suitable for the particular data. Combined with the relative simplicity of these functions, one expects that the processing speed could be increased. 

This paper is organized as follows. In section 2 we review some basic facts from the theory of Weyl groups. In section 3 the sign homomorphisms and their kernels are introduced. Sections 4 and 5 present in detail the even orbit function and the mixed even orbit function. The product of the $E-$functions is studied in the section 6.

\section{Weyl groups and corresponding fundamental domains}

\subsection{Weyl group and affine Weyl group}\

The ordered set of simple roots $\Delta=(\al_1,\al_2)$ of a simple Lie algebra of rank $2$ is a collection of $2$ vectors spanning a real $2-$dimensional Euclidean space $\R^2$ \cite{Hum,Kane}. The simple roots of $\Delta$ form a basis of $\R^2$ satisfying certain specific conditions. These roots are specified by their lengths and the angle between them. Equivalently, the root system $\Delta$ can be determined either by the Coxeter--Dynkin diagram (and the corresponding Coxeter matrix $M$) or the Cartan matrix $C$ of the simple Lie algebra.  
The coroots $\alpha_i^\vee$ are defined as $\alpha_i^\vee= 2\alpha_i/\left\langle \alpha_i, \alpha_i \right\rangle,\,i=1,2$.
In addition to the $\alpha-$basis of simple roots, we define the weight $\omega-$basis by
$$\left\langle \alpha_i^\vee,\omega_j\right\rangle=\delta_{ij}\,,\quad i,j\in\{1,2\}. $$ The fourth basis, called the coweight $\omega^\vee-$basis, is given by $\om_i^\vee= 2\om_i/\left\langle \alpha_i, \alpha_i \right\rangle,\,i=1,2.$

We define the root lattice $Q$ as the set of all integer linear combinations of simple roots
$$Q = \Z \alpha_1 + \Z \alpha_2 $$
and similarly we define the coroot $Q^\vee$ lattice by
$$Q^\vee = \Z \alpha^\vee_1 + \Z \alpha^\vee_2. $$
Moreover, we define the weight lattice and the coweight lattice  
$$P = \Z \om_1 + \Z \om_2,\q P^\vee = \Z \om^\vee_1 + \Z \om^\vee_2. $$

Some important subsets of the weight lattice $P$ are the cone of dominant weights $P^+$ and the cone of strictly dominant weights $P^{++}$:
\begin{equation*}
P^+=\Z^{\geq 0}\omega_1+\Z^{\geq 0}\omega_2\  \supset \  P^{++}=\N\omega_1+\N\omega_2.
\end{equation*}

The reflection $r_{\alpha}$, $\alpha \in \Delta$, which fixes the hyperplane orthogonal to $\alpha$ and passes through the origin can be explicitly written as
 $r_{\alpha}x=x-\langle \alpha,x\rangle\alpha^\vee $, where $x\in\R^2$.

Given a simple Lie algebra with the set of simple roots $\Delta=(\al_1,\al_2)$, the associated Weyl group $W$ is a finite group generated by reflections $r_i\equiv r_{\al_i},\,i=1,2$. The system of vectors $W\Delta$ ($W\Delta$ denotes $W$ acting on the simple roots $\Delta$) is the root system and contains the highest root $\xi \in W\Delta$. The affine reflection $r_0$ with respect to the highest root is given by
\begin{equation*}
r_0 x=r_\xi x + \frac{2\xi}{\sca{\xi}{\xi}}\,,\qquad
r_{\xi}x=x-\frac{2\sca{x}{\xi} }{\sca{\xi}{\xi}}\xi\,,\qquad x\in\R^2\,.
\end{equation*} 
By adding the affine reflection $r_0$ to the set of generators $\{r_1,r_2\}$ one obtains the affine Weyl group $W^{\mathrm{aff}}$. The affine Weyl group $W^{\mathrm{aff}}$ consists of transformations of $\R^2$ from $W$ and of shifts by vectors from the coroot lattice $Q^\vee$. In fact it holds that
$W^{\mathrm{aff}}= Q^\vee \rtimes W$.
The fundamental domain $F$ of the action of $W^{\mathrm{aff}}$ on $\R^2$ is a triangle with vertices $\left\{ 0, \frac{\om^{\vee}_1}{m_1},\frac{\om^{\vee}_2}{m_2} \right\}$, where $m_1,m_2$ are the coefficients of the highest root $\xi$ in $\al-$basis, $\xi=m_1\al_1 +m_2 \al_2$.

The set of dual roots $\Delta^\vee=(\al^\vee_1,\al^\vee_2)$ also generates a Weyl group $W$. The system of vectors $W\Delta^\vee$ is a root system and contains the highest dual root $\eta \in W\Delta^\vee$. The dual affine reflection $r^\vee_0$ with respect to the highest dual root is given by
\begin{equation*}
r^\vee_0 x=r_\eta x + \frac{2\eta}{\sca{\eta}{\eta}}\,,\qquad
r_{\eta}x=x-\frac{2\sca{x}{\eta} }{\sca{\eta}{\eta}}\eta\,,\qquad x\in\R^2\,.
\end{equation*} 
By adding the dual affine reflection $r^\vee_0$ to the set of generators $\{r_1,r_2\}$ 
one obtains the dual affine Weyl group $\widehat{W}^{\mathrm{aff}}$, see \cite{HP1}. The dual affine Weyl group $\widehat{W}^{\mathrm{aff}}$ consists of transformations of $\R^2$ from $W$ and of shifts by vectors from the root lattice $Q$; it holds that
$\widehat{W}^{\mathrm{aff}}= Q \rtimes W$.
The dual fundamental domain $F^\vee$ of the action of $\widehat{W}^{\mathrm{aff}}$ on $\R^2$ is a triangle with vertices $\left\{ 0, \frac{\om_1}{m^\vee_1},\frac{\om_2}{m^\vee_2} \right\}$, where $m^\vee_1,m^\vee_2$ are the coefficients of the highest dual root $\eta$ in $\al^\vee-$basis, $\eta=m^\vee_1\al^\vee_1 +m^\vee_2 \al^\vee_2$.

\subsection{Fundamental domains}\

It is advantageous to distinguish explicitly the different root lengths in $C_2$ and $G_2$. Instead of the generic set of simple roots of rank $2$ of the form 
$\Delta=(\al_1,\al_2)$, we use the following notation
\begin{align}
\Delta (C_2)&=(\al_s,\al_l)\label{rootC2}\\
\Delta (G_2)&=(\al_l,\al_s).\label{rootG2}
\end{align}
The symbol $\al_s$ denotes the short simple root. For $C_2$ we use the standard normalization $\sca{\al_s}{\al_s}=1$ and for $G_2$ we have $\sca{\al_s}{\al_s}=2/3$.
The symbol $\al_l$ denotes the long simple root. For $C_2$ and $G_2$ we have $\sca{\al_l}{\al_l}=2$. The angle between $\al_s$ and $\al_l$ is $3\pi /4$ for $C_2$ and $5\pi /6$ for $G_2$. The highest root $\xi$ is equal to $2\al_s+\al_l$ for $C_2$ and to $2\al_l+3\al_s$ for $G_2$.
The ordering of the roots in bases $\Delta(C_2)$ and $\Delta(G_2)$ in (\ref{rootC2}), (\ref{rootG2}) is in accordance with the standard convention. Furthermore, we have the corresponding coroot $\alpha_s^\vee= 2\alpha_s/\left\langle \alpha_s, \alpha_s \right\rangle$, the weight $\om_s$ satisfying $\sca{\al^\vee_s}{\om_s}=1$, the coweight $\om^\vee_s=2\om_s/\left\langle \alpha_s, \alpha_s \right\rangle$ and the corresponding reflection $r_s$ (which we call short reflection). We define $\al^\vee_l$, $\om_l$, $\om^\vee_l$ and $r_l$ (long reflection) analogously. Therefore, the fundamental domains $F$ have the following explicit form
\begin{align*}
F(C_2)&=\left\lbrace a\o_s^\vee+b\o_l^\vee\, |\, a,\,b\ge 0,\, 2a+b\le 1\right\rbrace \\
F(G_2)&=\left\lbrace a\o_l^\vee+b\o_s^\vee\, |\, a,\, b\ge 0,\, 2a+3b\le 1\right\rbrace.
\end{align*} 

\begin{figure}
\resizebox{7.5cm}{!}{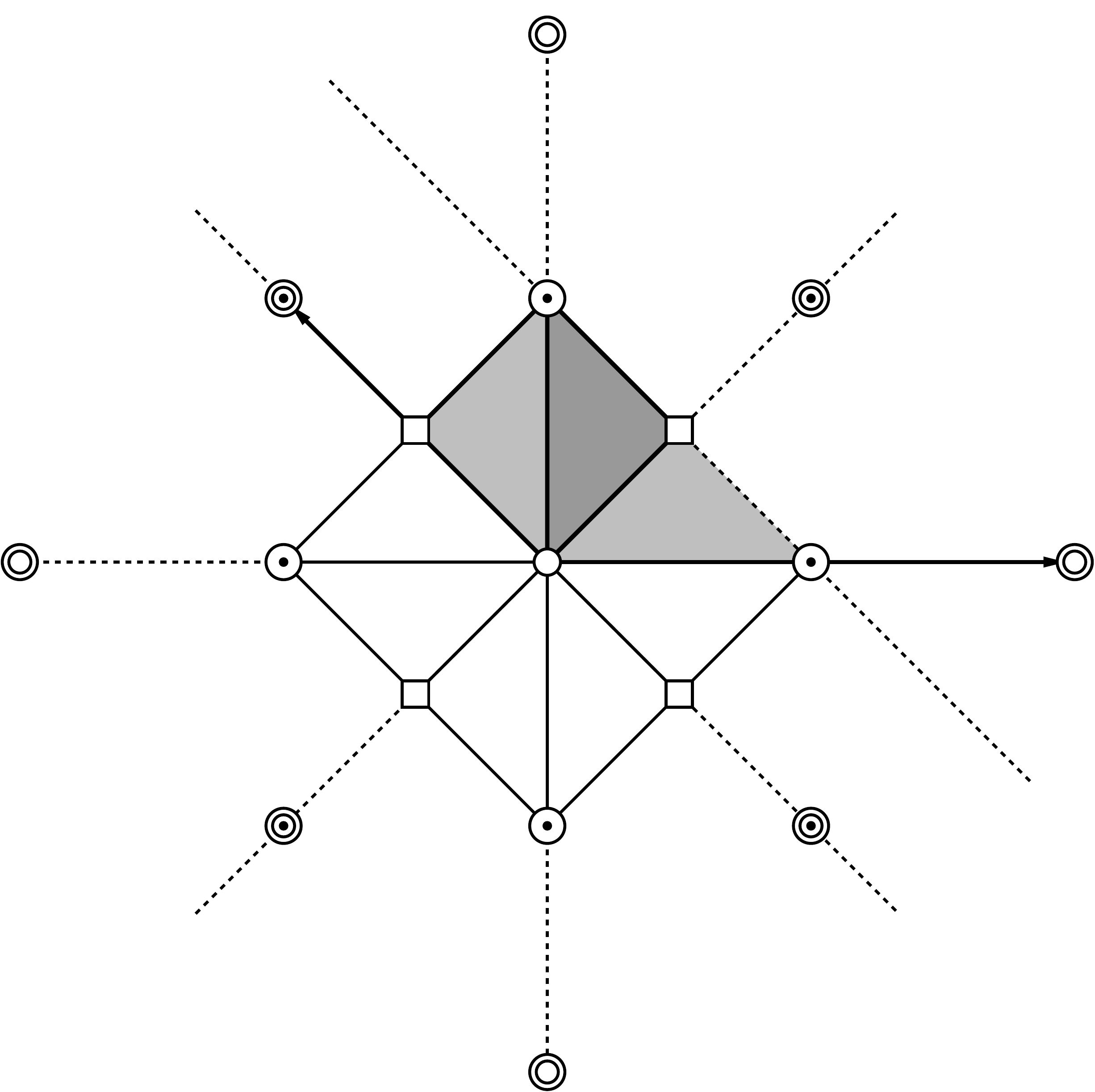}\hspace{10pt}
\resizebox{7.8cm}{!}{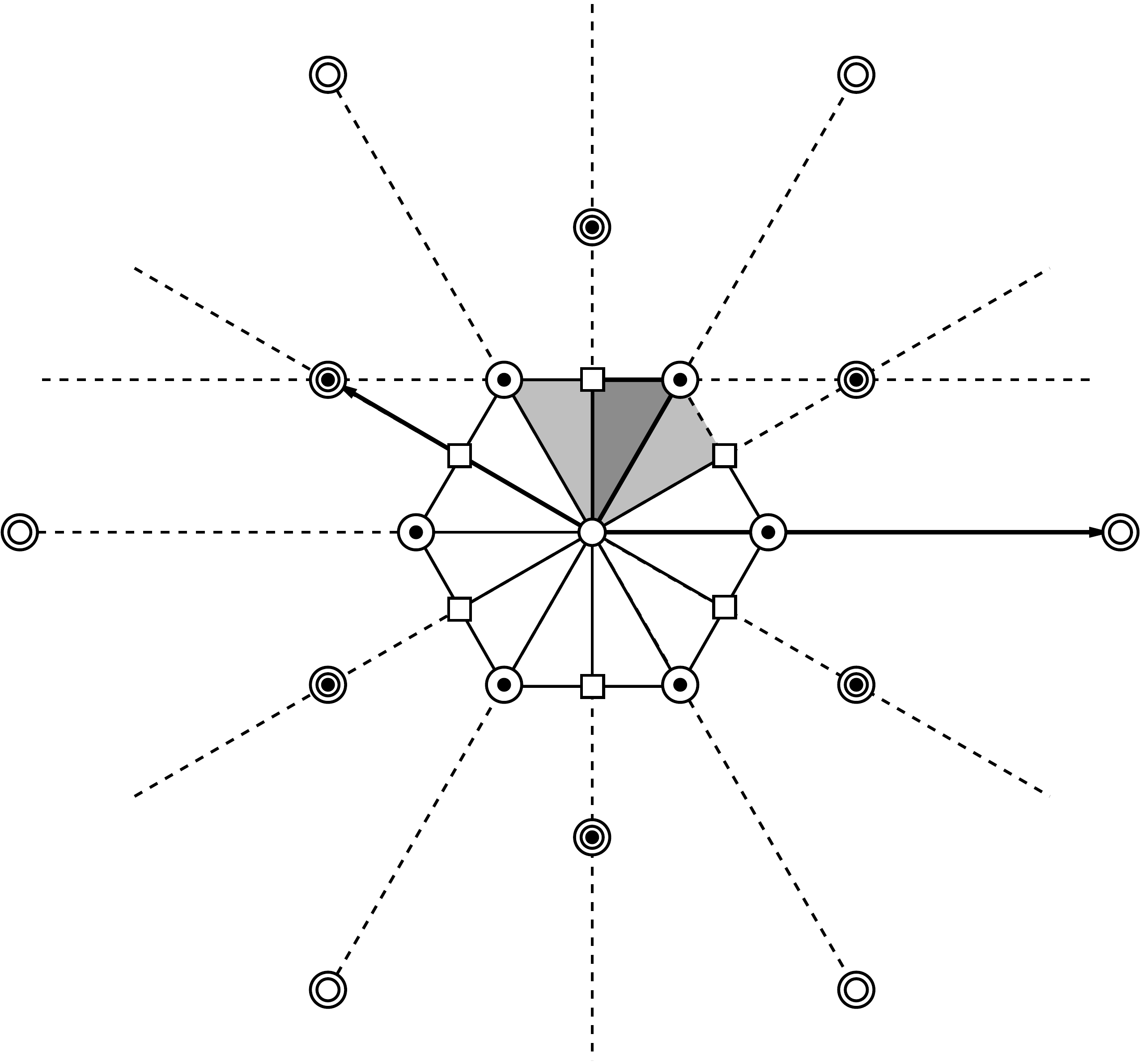}
\caption{The fundamental domains $F$ and the root systems of $C_2$ and $G_2$; the circles with a small dot inscribed depict the roots of the root system $W\Delta$ and the circles with a smaller circle inside them depict the elements of the dual root system $W\Delta^\vee$.}\label{fund}
\end{figure}

We also denote by $X_s$, $X_l$ the lines ('mirrors') which are stabilized by $r_s$, $r_l$, i.e. orthogonal to $\al_s$ and $\al_l$, respectively: 
\begin{equation*}
X_s=\set{x\in \R^2}{\sca{x}{\al_s}=0},\q X_l=\set{x\in \R^2}{\sca{x}{\al_l}=0}.
\end{equation*}
The lines which are stabilized with respect to the affine reflections $r_0$, $r^\vee_0$  are denoted by $X_0$, $X^\vee_0$, i.e. 
\begin{equation*}
X_0=\set{x\in \R^2}{\sca{x}{\xi}=1},\q X^\vee_0=\set{x\in \R^2}{\sca{x}{\eta}=1}.
\end{equation*}
The segments of the lines $X_s$, $X_l$ and $X_0$ which lie in the fundamental domain $F$ are denoted by $Y_s$, $Y_l$ and $Y_0$, respectively:
\begin{equation*}
Y_s=X_s\cap F, \q Y_l=X_l\cap F, \q Y_0=X_0\cap F.
\end{equation*}
Analogously, we define
\begin{equation*}
Y^\vee_s=X_s\cap F^\vee, \q Y^\vee_l=X_l\cap F^\vee, \q Y^\vee_0=X^\vee_0\cap F^\vee.
\end{equation*}
We distinguish the weights from the positive weight lattice $P^+$ which lie on the mirrors $X_s$, and $X_l$:
\begin{equation*}
P^s=X_s\cap P^+, \q P^l=X_l\cap P^+.
\end{equation*}
The fundamental domains $F$ together with the root systems of $C_2$ and $G_2$ are depicted in Figure~\ref{fund}.

\section{Homomorphisms and orbits}\

\subsection{Sign homomorphisms}\label{homo}\

The Weyl group $W$ can also be abstractly defined by the following presentation 
\begin{equation}\label{presentation}r_i^2=1,\quad (r_ir_j)^{m_{ij}}=1,\quad i,j=1,2\end{equation}
where integers $m_{ij}$ denote elements of the Coxeter matrix. 

Crucial for us are certain 'sign' homomorphisms $\sigma:W\rightarrow\{\pm1 \}.$ An admissible $\sigma$ has to satisfy the presentation condition \eqref{presentation}
\begin{equation}\label{admit}
\sigma(r_i)^2=1,\quad (\sigma(r_i)\sigma(r_j))^{m_{ij}}=1,\quad i,j=1,2.
\end{equation}
There are two obvious choices $\id,\, \sigma^e$ of such sign homomorphisms which are defined for any $w\in W$
\begin{align}
\id(w)&=1 \label{homid} \\
\sigma^e(w)&=\det w. \label{home}
\end{align}
It turns out that for root systems with two different lengths of roots there are two other choices available \cite{MMP}. 

The Lie algebras $C_2$ and $G_2$ are the only simple Lie algebras of rank $2$ which have two different lengths of roots. As such, they admit other sign homomorphisms (and corresponding special functions) besides the standard choices (\ref{homid}), (\ref{home}). 
The Coxeter matrices $M$ of $C_2$ and $G_2$ are the following:
\begin{equation}\label{Coxmat}
M(C_2)=\begin{pmatrix} 1 &4\\4&1\end{pmatrix}\,,\qquad M(G_2)=\,\begin{pmatrix} 1 &6\\6&1\end{pmatrix}\,.
\end{equation}
   
A sign homomorphism $\sigma :W \map\{\pm 1\} $ can be defined by prescribing its values on the generators $r_s$ and $r_l$ such that \eqref{admit} is satisfied. The two obvious choices $\sigma(r_s)=\sigma(r_l)=1$ and $\sigma^e(r_s)=\sigma^e(r_l)=-1$ lead to the standard homomorphisms $\id$ and $\sigma^e$. For the rank $2$ cases $C_2$ and $G_2$ the elements $m_{12}$ of the Coxeter matrices (\ref{Coxmat}) are even. Therefore, $\sigma(r_l)$ and $\sigma(r_s)$ can be independently $\pm1$ and \eqref{admit} is still satisfied. Consequently, there are two more sign homomorphisms, which we denote by $\sigma^s$ and $\sigma^l$:
\begin{align}
\sigma^s(r_s)&=-1\,,\quad\sigma^s(r_l)=1\,,\label{homs}\\
\sigma^l(r_l)&=-1\,,\quad\sigma^l(r_s)=1\,.\label{homl}
\end{align}
Every element $w$ from  a Weyl group $W$ can be written as a product of generators $w=r_{i_1}\ldots r_{i_k}$ where $r_{i_j}\in \{r_s,r_l\}$. Equivalently, we 
reformulate the definition (\ref{homs}), (\ref{homl}):

\begin{align*}
\sigma^s(w)&=\begin{cases} 1\quad \text {if there is an even number of short reflections $r_s$ in } w\\ -1\quad \text {if there is an odd number of short reflections $r_s$ in } w\,,\end{cases}\\
\sigma^l(w)&=\begin{cases} 1\quad \text {if there is an even number of long reflections $r_l$ in } w\\ -1\quad \text {if there is an odd number of long reflections $r_l$ in } w\,.\end{cases}\\
\end{align*}

\subsection{Orbits and stabilizers}\

Except for the trivial homomorphism $\id$, the kernels $\ker \sigma$ of the sign homomorphisms $\sigma$ form normal subgroups of index $2$ in $W$. We denote the kernels of $\sigma^e$, $\sigma^s$ and $\sigma^l$ by $W^e$, $W^s$ and $W^l$, respectively: 
\begin{equation*} 
W^e=\left\lbrace w\in W\,|\, \sigma^e(w)=1\right\rbrace ,  \q W^s=\left\lbrace w\in W\,|\, \sigma^s(w)=1\right\rbrace , \q W^l=\left\lbrace w\in W\,|\, \sigma^l(w)=1\right\rbrace .
\end{equation*}
Some general properties of the group  $W^e$ were derived in \cite{HP2}. 
Explicit knowledge of the orbits and stabilizers of $W^e$, $W^s$ and $W^l$ will be needed for description of orbit functions. Generic orbits of the action of $W^e$, $W^s$ and $W^l$ on $\R^2$ for the cases of $C_2$ and $G_2$ are shown in Figures \ref{orbitsC2} and \ref{orbitsG}.
\begin{figure}[ht]
\hspace{-40pt}
\resizebox{15cm}{!}{
\begin{picture}(440,140)
\put(0,120){$W^e$}\put(70,120){$(a,b)$}\put(55,115){\vector(-1,-3){15}}%
\put(0,60){$(a+2b,-a-b)$}\put(40,55){\vector(1,-3){15}}%
\put(60,0){$(-a,-b)$}\put(110,115){\vector(1,-3){15}}%
\put(90,60){$(-a-2b,a+b)$}\put(125,55){\vector(-1,-3){15}}%
\put(30,100){\begin{scriptsize}$r_lr_s$\end{scriptsize}}%
\put(120,100){\begin{scriptsize}$r_sr_l$\end{scriptsize}}%
\put(25,40){\begin{scriptsize}$r_lr_s$\end{scriptsize}}%
\put(125,40){\begin{scriptsize}$r_sr_l$\end{scriptsize}}%
\put(175,120){$W^s$}\put(234,120){$(a,b)$}\put(223,115){\vector(-1,-4){18}}%
\put(175,30){$(-a-2b,b)$}\put(208,25){\vector(1,-1){15}}%
\put(224,0){$(-a,-b)$}\put(268,115){\vector(1,-1){15}}%
\put(255,90){$(a+2b,-b)$}\put(285,85){\vector(-1,-4){18}}%
\put(207,13){\begin{scriptsize}$r_l$\end{scriptsize}}%
\put(280,110){\begin{scriptsize}$r_l$\end{scriptsize}}%
\put(193,100){\begin{scriptsize}$r_sr_lr_s$\end{scriptsize}}%
\put(288,70){\begin{scriptsize}$r_sr_lr_s$\end{scriptsize}}%
\put(330,120){$W^l$}\put(385,120){$(a,b)$}\put(375,115){\vector(-1,-1){15}}%
\put(330,90){$(-a,a+b)$}\put(360,85){\vector(1,-4){18}}%
\put(380,0){$(-a,-b)$}\put(420,115){\vector(1,-4){18}}%
\put(420,30){$(a,-a-b)$}\put(438,25){\vector(-1,-1){15}}%
\put(358,110){\begin{scriptsize}$r_s$\end{scriptsize}}%
\put(432,13){\begin{scriptsize}$r_s$\end{scriptsize}}%
\put(430,100){\begin{scriptsize}$r_lr_sr_l$\end{scriptsize}}%
\put(338,70){\begin{scriptsize}$r_lr_sr_l$\end{scriptsize}}%
\end{picture}}
\caption{Orbits of the actions of the groups $W^e$, $W^s$ and $W^l$ of $C_2$. The coordinates $(a,b)$ of the points in $\R^2$ are given in $\om-$basis. }\label{orbitsC2} 
\end{figure}
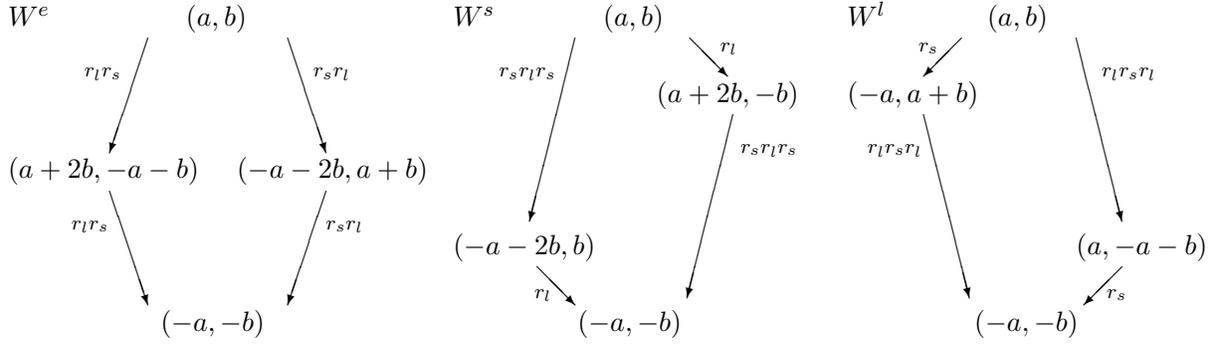

\begin{figure}[ht]
\resizebox{11cm}{!}{\begin{picture}(140,200)\put(0,180){$W^e$}
\put(70,180){$(a,b)$}\put(50,175){\vector(-1,-3){15}}%
\put(0,120){$(2a+b,-3a-b)$}\put(35,115){\vector(0,-1){45}}%
\put(0,60){$(a+b,-3a-2b)$}\put(35,55){\vector(1,-3){15}}%
\put(60,0){$(-a,-b)$}\put(115,175){\vector(1,-3){15}}%
\put(100,120){$(-a-b,3a+2b)$}\put(130,115){\vector(0,-1){45}}%
\put(100,60){$(-2a-b,3a+b)$}\put(130,55){\vector(-1,-3){15}}%
\put(25,160){\begin{scriptsize}$r_sr_l$\end{scriptsize}}%
\put(125,160){\begin{scriptsize}$r_lr_s$\end{scriptsize}}%
\put(15,100){\begin{scriptsize}$r_sr_l$\end{scriptsize}}%
\put(20,40){\begin{scriptsize}$r_sr_l$\end{scriptsize}}%
\put(135,100){\begin{scriptsize}$r_lr_s$\end{scriptsize}}%
\put(133,40){\begin{scriptsize}$r_lr_s$\end{scriptsize}}%
\end{picture}\hspace{60pt}
\begin{picture}(150,200)\put(0,180){$W^s$}
\put(75,180){$(a,b)$}\put(55,175){\vector(-1,-1){15}}%
\put(15,150){$(-a,3a+b)$}\put(40,145){\vector(0,-1){75}}%
\put(5,60){$(a+b,-3a-2b)$}\put(40,55){\vector(0,-1){15}}%
\put(15,30){$(-a-b,b)$}\put(120,175){\vector(1,-4){18}}%
\put(100,90){$(2a+b,-3a-2b)$}\put(138,85){\vector(0,-1){15}}%
\put(105,60){$(-2a-b,3a+b)$}
\put(40,170){\begin{scriptsize}$r_l$\end{scriptsize}}%
\put(130,165){\begin{scriptsize}$r_sr_lr_s$\end{scriptsize}}%
\put(15,135){\begin{scriptsize}$r_sr_lr_s$\end{scriptsize}}%
\put(30,48){\begin{scriptsize}$r_l$\end{scriptsize}}%
\put(142,78){\begin{scriptsize}$r_l$\end{scriptsize}}%
\end{picture}}\\ \resizebox{5.8cm}{!}{\hspace{40pt}\begin{picture}(140,200)\put(0,180){$W^l$}
\put(75,180){$(a,b)$}\put(58,175){\vector(-1,-4){18}}%
\put(0,90){$(-2a-b,3a+2b)$}\put(40,85){\vector(0,-1){15}}%
\put(5,60){$(a+b,-3a-2b)$}\put(120,175){\vector(1,-1){15}}%
\put(115,150){$(a+b,-b)$}\put(135,145){\vector(0,-1){75}}%
\put(105,60){$(-2a-b,3a+b)$}\put(135,55){\vector(0,-1){15}}%
\put(115,30){$(a,-3a-b)$}%
\put(30,165){\begin{scriptsize}$r_lr_sr_l$\end{scriptsize}}%
\put(130,170){\begin{scriptsize}$r_s$\end{scriptsize}}%
\put(30,78){\begin{scriptsize}$r_s$\end{scriptsize}}%
\put(140,135){\begin{scriptsize}$r_lr_sr_l$\end{scriptsize}}%
\put(140,48){\begin{scriptsize}$r_s$\end{scriptsize}}%
\end{picture}}
\caption{Orbits of the actions of the groups $W^e$, $W^s$ and $W^l$ of $G_2$. The coordinates $(a,b)$ of the points in $\R^2$ are given in $\om-$basis.}\label{orbitsG} 
\end{figure}
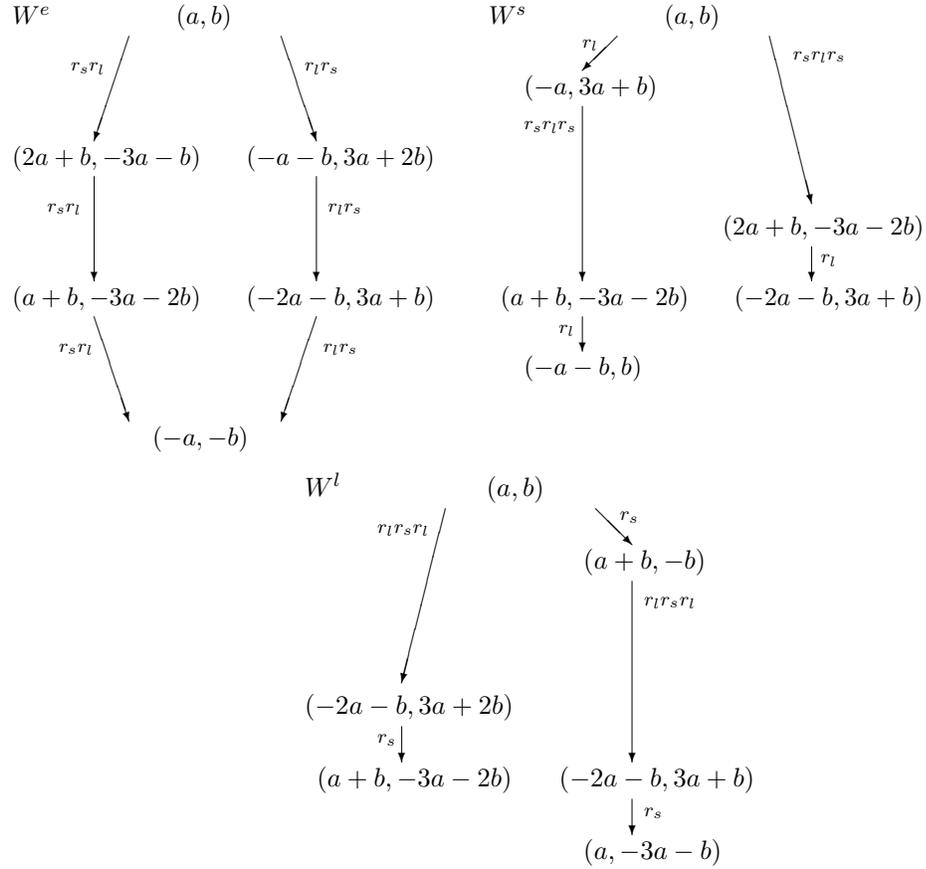

For any $\la\in \R^2$ we have the stabilizer $\mathrm{Stab}_{\ker\sigma} (\la)$ of $\la$
$$\mathrm{Stab}_{\ker\sigma} (\la)=\set{w\in \ker\sigma}{w\la=\la}$$
and we denote the orders of the stabilizers of $W^e$, $W^s$ and $W^l$ by $d^{e}_{\la}$, $d^{s}_{\la}$ and $d^{l}_{\la}$, respectively
\begin{equation}\label{dla}
d^{e}_{\la}\equiv  |\mathrm{Stab}_{W^e} (\la)|,\q d^{s}_{\la}\equiv  |\mathrm{Stab}_{W^s} (\la)|,\q d^{l}_{\la}\equiv  |\mathrm{Stab}_{W^l} (\la)|.
\end{equation}
For our purposes, it is sufficient to consider only values of $\la\in P^+$. If $\la \in r P^+$, for $r=r_s, r_l$, then the stabilizers of $\la$ and $r\la$ are conjugate and have the same order, i.e.
\begin{equation*}
d^{e}_{\la}=d^e_{r\la},\q d^{l}_{\la}=d^l_{r\la}, \q d^{s}_{\la}=d^s_{r\la},\q r=r_s,r_l.
\end{equation*}
The orders of stabilizers $d^{e}_{\la}$, $d^{s}_{\la}$ and $d^{l}_{\la}$ with $\la\in P^+$ are for the cases of $C_2$ and $G_2$ shown in Table~\ref{tabStab}.
{\small
\begin{table}[ht]
\begin{tabular}{|c|c|c|c|c|c|c|}
\hline\multirow{2}{*}{$\la\in P^{+}$} 
&\multicolumn{3}{|c|}{$C_2$}& \multicolumn{3}{|c|}{$G_2$}\\
\cline{2-7}
&$d^e_\la$ & $d^s_\la$ &$d^l_\la$&$d^e_\la$ &$d^s_\la$&$d^l_\la$\\
\hline\hline
$(a,b)$ &1&1&1&1&1&1\\
$(a,0)$ &1&2&1&1&1&2\\
$(0,b)$&1&1&2&1&2&1\\
$(0,0)$&4&4&4&6&6&6\\ \hline
\end{tabular}
\medskip
\caption{Orders of stabilizers of $\lambda\in P^+$ for $C_2$ and $G_2$, The coordinates $(a,b)$ are in $\om-$basis with $a\neq 0$, $b\neq 0$.}\label{tabStab}
\end{table}}

While calculating continuous orthogonality of various types of orbit functions, the number of elements in the Weyl group and the volume of the fundamental domain often appear. We denote the number $|W||F|$ by $K$ and we have (see e.g. \cite{HP1})
\begin{equation}\label{K}
K\equiv |W||F|= 
  \begin{cases}
    2 & \text{for } C_2 \\
      \sqrt{3}    & \text{for } G_2.
  \end{cases}
\end{equation}

\subsection{Orbits and stabilizers on the maximal torus}\

The orbits and stabilizers of the maximal torus will be needed for the discrete calculus of orbit functions. We choose some arbitrary natural number $M$ which will control the density of the grids appearing in this calculus \cite{HP1}. The discrete calculus of orbit functions is performed over the finite group $\frac{1}{M}P^{\vee}/Q^{\vee}$. The finite complement set of weights is taken as the quotient group $P/MQ$.  
For $x\in \R^2/Q^{\vee}$, we denote the orbit of a subgroup $\ker\sigma$  by
\begin{equation*}
(\ker\sigma) x=\set{wx\in \R^2/Q^{\vee} }{w\in \ker\sigma}.
\end{equation*}
The orders of the group stabilizers of $W^e$, $W^s$ and $W^l$ are denoted by $\ep^e$, $\ep^s$ and $\ep^l$:
\begin{equation}\label{stabx}
\ep^e(x)\equiv |W^e x |,\q \ep^s(x)\equiv |W^s x |,\q \ep^l(x)\equiv |W^l x |.
\end{equation}
For practical purposes it is sufficient to determine the sizes of these orbits for the finite set $$F_M\equiv  \frac{1}{M}P^{\vee}/Q^{\vee} \cap F .$$ For a general review of these sets $F_M$ see \cite{HP1}. 
If $x \in r F_M$, for $r=r_s,r_l$, then the orbits of $x$ and $r x$ have identical size, i.e. for $x\in F_M$
\begin{equation*}
\ep^{e}(x)=\ep^e(rx)\q\ep^{s}(x)=\ep^s(rx), \q \ep^{l}(x)=\ep^l(rx),\q r=r_s,r_l.
\end{equation*}
For our cases $C_2$ and $G_2$ the sets $F_M$ are of the following explicit form
\begin{align}
F_M(C_2)&=\setb{ \frac{a}{M}\o_s^\vee+\frac{b}{M }\o_l^\vee}{ a, b,c \in \Z^{\ge 0}, c+2a+b= M} \label{FMC2} \\
F_M(G_2)&=\setb{ \frac{a}{M}\o_l^\vee+\frac{b}{M }\o_s^\vee}{ a, b,c \in \Z^{\ge 0}, c+2a+3b= M}. \label{FMG2}
\end{align}
The coefficients $\ep^e(x)$, $\ep^s(x)$ and $\ep^l(x)$ for the cases $C_2$ and $G_2$ are listed in Table~\ref{tabOrb}.
{\small\begin{table}[ht]
\begin{tabular}{|c|c|c|c|c|c|c|}
\hline\multirow{2}{*}{$x\in F_M$} 
&\multicolumn{3}{|c|}{$C_2$}& \multicolumn{3}{|c|}{$G_2$}\\
\cline{2-7}
&$\ep^e(x)$ & $\ep^s(x)$ &$\ep^l(x)$&$\ep^e(x)$ &$\ep^s(x)$&$\ep^l(x)$\\
\hline\hline
$[c,a,b]$ &4&4&4&6&6&6\\
$[0,a,b]$ &4&2&4&6&3&6\\
$[c,0,b]$ &4&4&2&6&3&6\\
$[c,a,0]$ &4&2&4&6&6&3\\ 
$[0,0,b]$ &1&1&1&2&1&2 \\
$[0,a,0]$ &2&1&2&3&3&3 \\
$[c,0,0]$ &1&1&1&1&1&1 \\
\hline
\end{tabular}\hspace{10pt}
\begin{tabular}{|c|c|c|c|c|c|c|}
\hline\multirow{2}{*}{$\la\in \Lambda_M$} 
&\multicolumn{3}{|c|}{$C_2$}& \multicolumn{3}{|c|}{$G_2$}\\
\cline{2-7}
&$h^{e\vee}_{\la}$ & $h^{s\vee}_{\la}$ &$h^{l\vee}_{\la}$&$h^{e\vee}_{\la}$ &$h^{s\vee}_{\la}$&$h^{l\vee}_{\la}$\\
\hline\hline
$[c,a,b]$ &1&1&1&1&1&1\\
$[0,a,b]$ &1&1&2&1&1&2\\
$[c,0,b]$ &1&1&2&1&2&1\\
$[c,a,0]$ &1&2&1&1&1&2\\ 
$[0,0,b]$ &2&2&4&2&2&2 \\
$[0,a,0]$ &4&4&4&3&3&6 \\
$[c,0,0]$ &4&4&4&6&6&6 \\
\hline
\end{tabular}\medskip
\caption{Orders of orbits of $x\in F_M$ and stabilizers of $\la\in \Lambda_M$ for the cases $C_2$ and $G_2$. The coordinates $[c,a,b]$ of $x\in F_M(C_2)$, $x\in F_M(G_2)$ are as in (\ref{FMC2}), (\ref{FMG2}), respectively. The coordinates $[c,a,b]$ of $\la\in \Lambda_M(C_2)$ and $\la\in \Lambda_M(G_2)$ are taken from (\ref{LMC2}), (\ref{LMG2}), respectively. It is assumed that $a,b,c\neq 0$. }\label{tabOrb}
\end{table}}

For any $\la\in P/MQ$, we denote the stabilizer $\mathrm{Stab}^{\vee}_{\ker\sigma} (\la)$ of $\la$ by
$$\mathrm{Stab}^{\vee}_{\ker\sigma} (\la)=\set{w\in \ker\sigma}{w\la=\la}.$$
The corresponding orders of the stabilizers of the action of $W^e$, $W^s$ and $W^l$ on $P/MQ$ are denoted by $h^{e\vee}_{\la}$, $h^{s\vee}_{\la}$ and $h^{l\vee}_{\la}$, respectively
\begin{equation}\label{hla}
h^{e\vee}_{\la}\equiv |\mathrm{Stab}^{\vee}_{W^e} (\la)|,\q h^{s\vee}_{\la}\equiv |\mathrm{Stab}^{\vee}_{W^s} (\la)|,\q h^{l\vee}_{\la}\equiv |\mathrm{Stab}^{\vee}_{W^l} (\la)|.
\end{equation}
For practical purposes it is sufficient to determine the sizes of these stabilizers for the finite set $$\Lambda_M\equiv P/MQ \cap MF^\vee .$$ For a general review of the sets $\Lambda_M$ see \cite{HP1}. 
If $\la \in r \Lambda_M$, for $r=r_s,r_l$, then the stabilizers of $\la$ and $r \la$ are conjugate and have identical size, i.e. for $\la\in \Lambda_M$ we have
\begin{equation*}
h^{e\vee}_{\la}=h^{e\vee}_{r\la},\q h^{s\vee}_{\la}=h^{s\vee}_{r\la}, \q h^{l\vee}_{\la}=h^{l\vee}_{r\la},\q r=r_s,r_l.
\end{equation*}
For $C_2$ and $G_2$, the sets $\Lambda_M$ are of the following explicit form
\begin{align}
\Lambda_M(C_2)&=\setb{ a\o_s+b\o_l}{ a, b,c \in \Z^{\ge 0}, c+a+2b= M} \label{LMC2} \\
\Lambda_M(G_2)&=\setb{ a\o_l+b\o_s}{ a, b,c \in \Z^{\ge 0}, c+3a+2b= M} \label{LMG2}.
\end{align}
The coefficients $h^{e\vee}_{\la}$, $h^{s\vee}_{\la}$ and $h^{l\vee}_{\la}$ for the cases $C_2$ and $G_2$ are listed in Table~\ref{tabOrb}.

While calculating discrete orthogonality of various types of orbit functions, the number of elements of the Weyl group and the determinant of the Cartan matrix $\det C$ often appear. We denote the number $|W|\det C/2$ by $k$ and we have (see e.g. \cite{HP1})
\begin{equation}\label{k}
k\equiv \frac{\abs{W}\det C}{2}= 
  \begin{cases}
    8 & \text{for } C_2 \\
    6 & \text{for } G_2.
  \end{cases}
\end{equation}

\section{Even orbit functions}\

Any sign homomorphism $\sigma :W \map \{\pm 1\}$ determines, in general, a complex orbit function $\psi^\sigma_\la:\R^2\map \C$ parametrized by $\la\in P$: 
\begin{equation}\label{genorb}
\psi^\sigma_\la(x)=\sum_{w\in W}\sigma (w)\, e^{2 \pi i \sca{ w\la}{x}},\q x\in \R^2.
\end{equation}
For the choice (\ref{homid}) of $\id$ in (\ref{genorb}), we obtain the so-called $C-$functions \cite{KP2}. The $C-$functions $\psi^\id_\la$ were studied in detail for all rank two cases in \cite{PZ1,PZ2}. For the choice (\ref{homid}) of $\sigma^e$, we obtain the well-known $S-$functions \cite{KP3}.
The $S-$functions $\psi^{\sigma^e}_\la$ resulting from homomorphism $\sigma^e$ and (\ref{genorb}) were described for all rank two cases in \cite{PZ3}. The remaining two options of homomorphisms (\ref{homs}), (\ref{homl}) and corresponding functions $\psi^{\sigma^s}_\la$, $\psi^{\sigma^l}_\la$, called $S^l-$ and $S^s-$functions \cite{MMP}, were studied in detail for $G_2$ only recently in \cite{Sz}.

The kernels $\ker \sigma$ of the sign homomorphisms $\sigma$, $\sigma\neq \id$, give rise to another type of 'even' orbit functions. They are complex functions $\varPsi^\sigma_\la:\R^2\map \C$ parametrized by $\la\in P$,  
\begin{equation}\label{genorbeven}
\varPsi^\sigma_\la(x)=\sum_{w\in \ker \sigma } e^{2 \pi i \sca{ w\la}{x}},\q x\in \R^2.
\end{equation} 
These 'even' orbit functions are invariant with respect to the action of $w\in \ker \sigma$ 
\begin{align}
\varPsi^\sigma_\la(wx)&=\varPsi^\sigma_\la (x) \label{geninveven1}\\
\varPsi^\sigma_{w\la}(x)&=\varPsi^\sigma_\la (x)\label{geninveven2}
\end{align}
and the invariance with respect to shifts from $Q^\vee$ also holds
\begin{equation}\label{genshifteven}
\varPsi^\sigma_\la(x+q^\vee )=\varPsi^\sigma_\la (x),\q q^\vee \in Q^\vee.
\end{equation}
In the following sections we investigate all the possible choices of sign homomorphisms and the resulting functions. We also discuss basic properties of those functions, including continuous and discrete orthogonality and corresponding transforms. It turns out that for all cases similar orthogonality relations to those in \cite{HP2,MP1} hold. 

\subsection{$\Xi^{e+}-$functions}\

For the choice $\sigma^e$, we obtain the so-called $E-$functions \cite{KP1}. Here we denote these functions  by the symbol $\Xi^{e+}_\la\equiv \varPsi^{\sigma^e}_\la$ and the corresponding kernel is $ W^e$. The invariance (\ref{geninveven1}), (\ref{genshifteven}) with respect to $Q^\vee \rtimes W^e$ allows us to consider $\Xi^{e+}_\la$ only on the domain
\begin{equation*}
F^{e+}= F\cup r F^\circ
\end{equation*} 
where $F^\circ$ denotes the interior of $F$ and $r$ is an arbitrary generating reflection $r\in \{r_s,r_l\}$. Similarly, the invariance (\ref{geninveven2}) restricts $\la\in P$ to the set
\begin{equation*}
P_{e+}= P^{+}\cup r P^{++}.
\end{equation*} 
Thus, we have
\begin{equation*}
\Xi^{e+}_\la(x)=\sum_{w\in W^e } e^{2 \pi i \sca{ w\la}{x}},\q x\in F^{e+},\, \la \in P_{e+}.
\end{equation*} 
These $E-$functions $\Xi^{e+}_\la$ were studied for all rank two cases in detail in \cite{PK}.

\subsection{$\Xi^{s+}-$functions}\

We discuss the resulting functions $\varPsi^{\sigma^s}_\la$ which correspond to the sign homomorphism $\sigma^s$. 
We denote these functions by $\Xi^{s+}_\la\equiv \varPsi^{\sigma^s}_\la$ and the corresponding kernel is $ W^s$. The invariance (\ref{geninveven1}), (\ref{genshifteven}) with respect to $Q^\vee \rtimes W^s$ allows us to consider $\Xi^{s+}_\la$ only on the domain
\begin{equation*}
F^{s+}= F\cup r_s (F\setminus Y_s).
\end{equation*} 
Similarly, the invariance (\ref{geninveven2}) restricts $\la\in P$ to the set
\begin{equation*}
P_{s+}= P^{+}\cup r_s (P^{+}\setminus P^s).
\end{equation*} 
Thus, we have
\begin{equation*}
\Xi^{s+}_\la(x)=\sum_{w\in W^s } e^{2 \pi i \sca{ w\la}{x}},\q x\in F^{s+},\, \la \in P_{s+}.
\end{equation*} 

\subsubsection{Continuous orthogonality and $\Xi^{s+}-$transforms}
For any two weights $\la,\la'\in P_{s+}$ the
corresponding $\Xi^{s+}-$functions are orthogonal on $F^{s+}$
\begin{equation}\label{s+corthog}
\int_{{F}^{s+}}\Xi^{s+}_{\lambda}(x)\overline{\Xi^{s+}_{\lambda'}(x)}\,dx=
K\,d^{s}_{\la}\,\delta_{\lambda\lambda'} \end{equation}
where $d^{s}_{\la}$, $K$ are given by (\ref{dla}), (\ref{K}), respectively. The $\Xi^{s+}-$functions determine symmetrized Fourier series
expansions,
\begin{equation*}
f(x)=\sum_{\la\in P_{s+}}c^{s+}_{\la}\Xi^{s+}_{\la}(x),\quad {\mathrm{
where}}\
c^{s+}_{\la}=\frac{1}{K d^{s}_{\la}}\int_{F^{s+}}f(x)\overline{\Xi^{s+}_{\la}(x)}\,dx.
\end{equation*}

\subsubsection{Discrete orthogonality and discrete $\Xi^{s+}-$transforms}\ The finite set of points is given by $$F_M^{s+}=\frac{1}{M}P^\vee / Q^\vee \cap F^{s+}.  $$
We define the corresponding finite set of weights as 
$$\Lambda^{s+}_M = P/MQ \cap MF^{s+\vee}$$
where $$F^{s+\vee}=F^\vee \cup r_s (F^\vee \setminus (Y_s^\vee \cup Y_0^\vee)).$$
Then, for $\la,\la' \in\Lambda^{s+}_M$, the following discrete orthogonality relations hold:
\begin{equation}\label{s+dortho}
 \sum_{x\in F^{s+}_M}\ep^s(x) \Xi^{s+}_\la(x)\overline{\Xi^{s+}_{\la'}(x)}=kM^2 h^{s\vee}_\la \delta_{\la\la'}
\end{equation}
where $h^{s\vee}_{\la}$, $k$ are given by (\ref{hla}), (\ref{k}), respectively. The discrete symmetrized $\Xi^{s+}-$function expansion is given by                    \begin{equation*}
f(x)=\sum_{\la\in \Lambda_M^{s+}}c^{s+}_{\la}\Xi^{s+}_{\la}(x),\quad {\mathrm{
where}}\
c^{s+}_{\la}=\frac{1}{k M^2 h^{s\vee}_\la}\sum_{x\in F^{s+}_M}\ep^{s}(x) f(x)\overline{\Xi^{s+}_{\la}(x)}.
\end{equation*}

\subsubsection{$\Xi^{s+}-$functions of $C_2$}
\begin{figure}
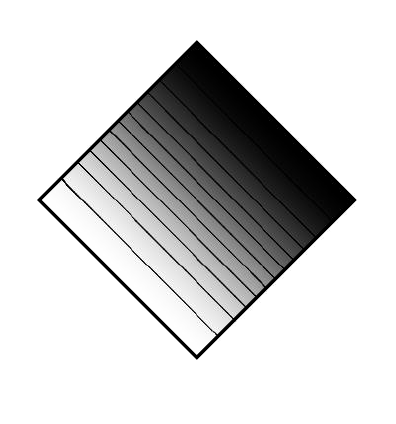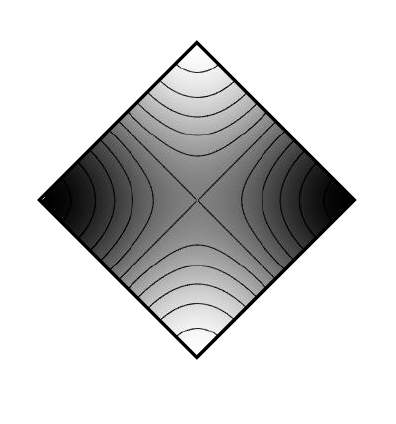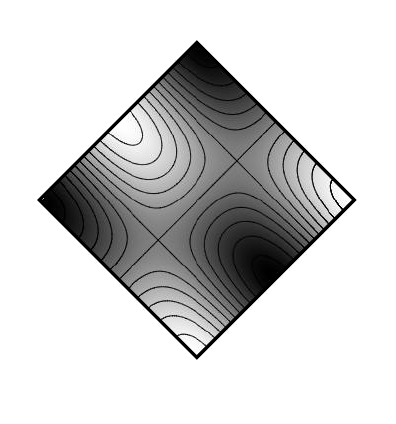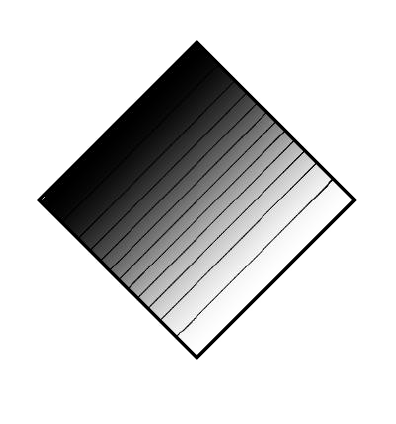
\caption{The contour plots of $\Xi^{s+}-$functions of $C_2$ over the fundamental domain $F^{s+}(C_2)$.}\label{EsC2}
\end{figure} 
For a point with coordinates in $\alpha^\vee$-basis $(x,y)$ we have the following explicit form of $\Xi^{s+}-$functions of $C_2$:
$$
\Xi^{s+}_{(a,b)}(x,y)=2\left\lbrace\cos(2\pi(ax+by))+\cos(2\pi((a+2b)x-by))\right\rbrace.
$$
The fundamental domain $F^{s+}$ is of the form
$$F^{s+}(C_2) =\left\lbrace x\o_s^\vee+y\o_l^\vee\, |\, x,\, y\ge 0, 2x+y\le 1\right\rbrace 
\cup \left\lbrace -x\o_s^\vee+(2x+y)\o_l^\vee\, |\,x> 0, y\ge 0, 2x+y\le 1 \right\rbrace
$$ 
and the lattice of weights $P_{s+}$ is given by
$$P_{s+}(C_2) =\left\lbrace a\o_s+b\o_l\, |\,a,\, b\in \Z^{\ge0}\right\rbrace 
\cup \left\lbrace-a\o_s+(a+b)\o_l\, |\, a\in \N, b\in \Z^{\ge0}\right\rbrace .
$$ 
The contour plots of some lowest $\Xi^{s+}-$functions of $C_2$ are given in Figure \ref{EsC2}.
The coefficients $d_\la^s$ of continuous orthogonality relations (\ref{s+corthog}) are given in Table \ref{tabStab}.

The discrete grid $F_M^{s+}$ is given by
\begin{align*}F_M^{s+}(C_2) =&\left\lbrace\dfrac{a}{M}\o_s^\vee+\dfrac{b}{M}\o_l^\vee \, |\,c,\,a,\, b\in \Z^{\ge0},\, c+2a+b= M\right\rbrace\\ 
&\cup \left\lbrace-\dfrac{a}{M}\o_s^\vee+\dfrac{2a+b}{M}\o_l^\vee \, |\,a\in \N, c,\,b\in \Z^{\ge0},\, c+2a+b= M\right\rbrace
\end{align*}
and the corresponding finite set of weights has the form
\begin{align*}\Lambda_M^{s+}(C_2) =&\left\lbrace a\o_s+b\o_l\, |\, c,\,a,\,b \in \Z^{\ge0},\, c+a+2b= M\right\rbrace\\ 
&\cup \left\lbrace-a\o_s+(a+b)\o_l\, |\, a,\,c\in \N,\,b\in \Z^{\ge0},\,c+ a+2b= M\right\rbrace.
\end{align*}
The coefficients $\ep^{s}(x)$ and $h_\la^{s\vee }$ of discrete orthogonality relations (\ref{s+dortho}) are given in Table \ref{tabOrb}. 

\subsubsection{$\Xi^{s+}-$functions of $G_2$}
\begin{figure}
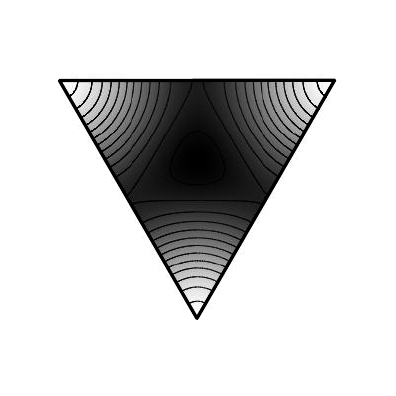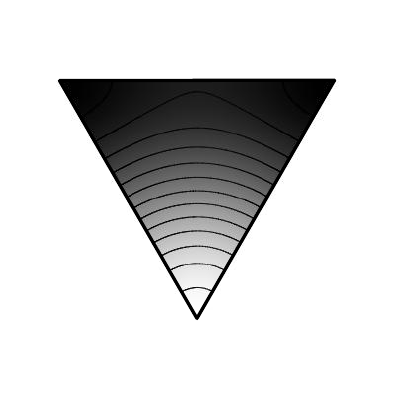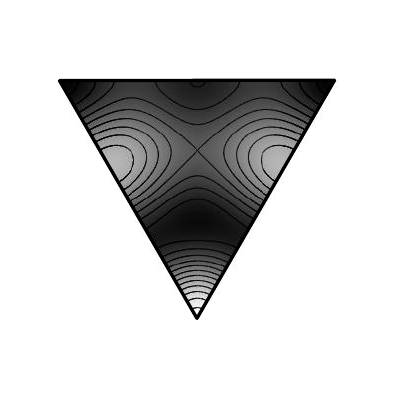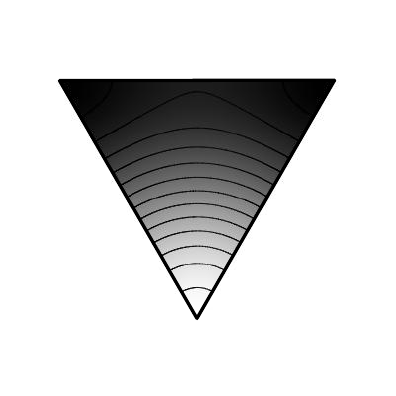\\ \vspace{-20pt}
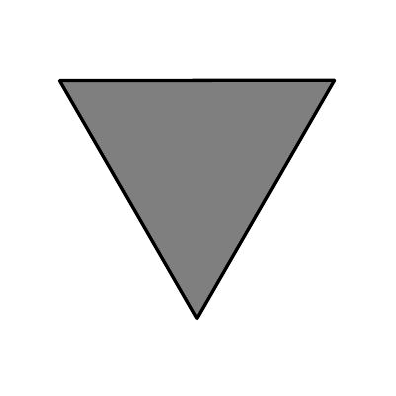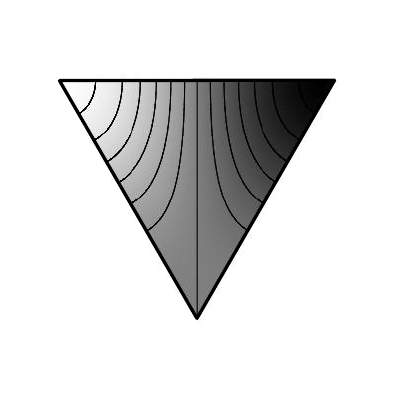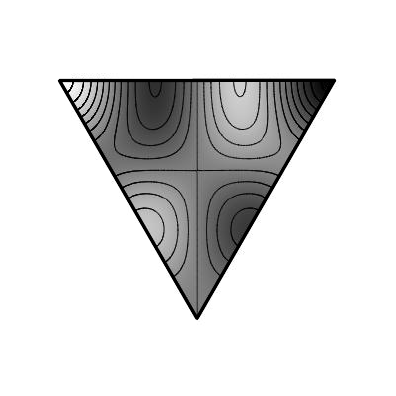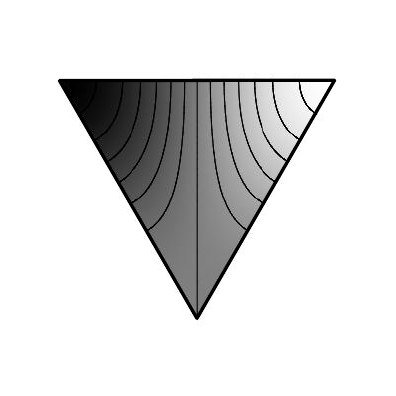
\caption{The contour plots of $\Xi^{s+}-$functions of $G_2$ over the fundamental domain $F^{s+}(G_2)$.}\label{EsG2}
\end{figure}

For a point with coordinates in $\alpha^\vee$-basis $(x,y)$ we have the following explicit form of $\Xi^{s+}-$functions of $G_2$:
\begin{align*}
\Xi^{s+}_{(a,b)}(x,y)=&e^{2\pi i(ax+by)}+ e^{2\pi i (-ax+(3a+b)y)}+ e^{2\pi i ((2a+b)x-(3a+2b)y)}\\ &+ e^{2\pi i ((a+b)x-(3a+2b)y)}+ e^{2\pi i (-(2a+b)x+(3a+b)y)}+ e^{2\pi i (-(a+b)x+by)}.
\end{align*}
The fundamental domain $F^{s+}$ is of the form
$$F^{s+}(G_2)
=\left\lbrace x\o_l^\vee+y\o_s^\vee \, |\,x,\,y\ge0,2x+3y\le1\right\rbrace\cup \left\lbrace (x+3y)\o_l^\vee-y\o_s^\vee\, |\,x\ge0,y>0,2x+3y\le1\right\rbrace\\
$$ 
and the lattice of weights $P_{s+}$ is given by
$$P_{s+}(G_2) =\left\lbrace  a\o_l+b\o_s\, |\,a,\,b\in\Z^{\ge0}\right\rbrace
\cup \left\lbrace (a+b)\o_l-b\o_s \, |\,a\in\Z^{\ge0},b\in\N\right\rbrace .
$$ 
The contour plots of some lowest $\Xi^{s+}-$functions of $G_2$ are given in Figure \ref{EsG2}.
The coefficients $d_\la^s$ of continuous orthogonality relations (\ref{s+corthog}) are given in Table \ref{tabStab}.

The discrete grid $F_M^{s+}$ is given by
\begin{align*}F_M^{s+}(G_2) =&\left\lbrace \dfrac{a}{M}\o_l^\vee+\dfrac{b}{M}\o_s^\vee \,\mid\, a,\,b,\,c\in \Z^{\ge0},\, c+ 2a+3b= M\right\rbrace\\ 
&\cup \left\lbrace\dfrac{a+3b}{M}\o_l^\vee-\dfrac{b}{M}\o_s^\vee \, |\,c,\,a\in \Z^{\ge0}, b\in \N,\, c+2a+3b= M\right\rbrace
\end{align*}
and the corresponding finite set of weights has the form
\begin{align*}\Lambda_M^{s+}(G_2) =&\left\lbrace a\o_l+b\o_s\, |\,a, \, b,\, c\in \Z^{\ge0},\,c+ 3a+2b= M\right\rbrace\\ 
&\cup \left\lbrace (a+b)\o_l-b\o_s \, |\, a\in \Z^{\ge0},\, b,\,c\in \N,\,c+ 3a+2b= M\right\rbrace.
\end{align*}
The coefficients $\ep^{s}(x)$ and $h_\la^{s\vee }$ of discrete orthogonality relations (\ref{s+dortho}) are given in Table \ref{tabOrb}. 

\subsection{$\Xi^{l+}-$functions}\

We now focus on the functions $\varPsi^{\sigma^l}_\la$ which correspond to the sign homomorphism $\sigma^l$. 
We denote these functions by $\Xi^{l+}_\la\equiv \varPsi^{\sigma^l}_\la$ and the corresponding kernel is $ W^l$. The invariance (\ref{geninveven1}), (\ref{genshifteven}) with respect to $Q^\vee \rtimes W^l$ allows us to consider $\Xi^{l+}_\la$ only on the domain
\begin{equation*}
F^{l+}= F\cup r_l (F\setminus (Y_l\cup Y_0)).
\end{equation*} 
Similarly, the invariance (\ref{geninveven2}) restricts $\la\in P$ to the set
\begin{equation*}
P_{l+}= P^{+}\cup r_l (P^{+}\setminus P^l).
\end{equation*} 
Thus, we have
\begin{equation*}
\Xi^{l+}_\la(x)=\sum_{w\in W^l } e^{2 \pi i \sca{ w\la}{x}},\q x\in F^{l+},\, \la \in P_{l+}.
\end{equation*} 

\subsubsection{Continuous orthogonality and $\Xi^{l+}-$transforms}
For any two weights $\la,\la'\in P_{l+}$ the
corresponding $\Xi^{l+}-$functions are orthogonal on $F^{l+}$
\begin{equation}\label{l+corthog}
\int_{{F}^{l+}}\Xi^{l+}_{\lambda}(x)\overline{\Xi^{l+}_{\lambda'}(x)}\,dx=
K\,d^{l}_{\la}\,\delta_{\lambda\lambda'} \end{equation}
where $d^{l}_{\la}$, $K$ are given by (\ref{dla}), (\ref{K}), respectively. The $\Xi^{l+}-$functions determine symmetrized Fourier series
expansions,
\begin{equation*}
f(x)=\sum_{\la\in P_{l+}}c^{l+}_{\la}\Xi^{l+}_{\la}(x),\quad {\mathrm{
where}}\
c^{l+}_{\la}=\frac{1}{K d^{l}_{\la}}\int_{F^{l+}}f(x)\overline{\Xi^{l+}_{\la}(x)}\,dx.
\end{equation*}

\subsubsection{Discrete orthogonality and discrete $\Xi^{l+}-$transforms}\ The finite set of points is given by $$F_M^{l+}=\frac{1}{M}P^\vee / Q^\vee \cap F^{l+}.  $$
We define the corresponding finite set of weights as 
$$\Lambda^{l+}_M = P/MQ \cap MF^{l+\vee} $$
where $$F^{l+\vee}=F^\vee \cup r_l (F^\vee \setminus Y_l^\vee).$$
Then, for $\la,\la' \in\Lambda^{l+}_M$, the discrete orthogonality relations hold
\begin{equation}\label{l+dortho}
 \sum_{x\in F^{l+}_M}\ep^l(x) \Xi^{l+}_\la(x)\overline{\Xi^{l+}_{\la'}(x)}=kM^2 h^{l\vee}_\la \delta_{\la\la'}
\end{equation}
where $h^{l\vee}_{\la}$, $k$ are given by (\ref{hla}), (\ref{k}), respectively. The discrete symmetrized $\Xi^{l+}-$functions expansion is given by                    \begin{equation*}
f(x)=\sum_{\la\in \Lambda_M^{l+}}c^{l+}_{\la}\Xi^{l+}_{\la}(x),\quad {\mathrm{
where}}\
c^{l+}_{\la}=\frac{1}{k M^2 h^{l\vee}_\la}\sum_{x\in F^{l+}_M}\ep^{l}(x) f(x)\overline{\Xi^{l+}_{\la}(x)}.
\end{equation*}

\subsubsection{$\Xi^{l+}-$functions of $C_2$}
\begin{figure}
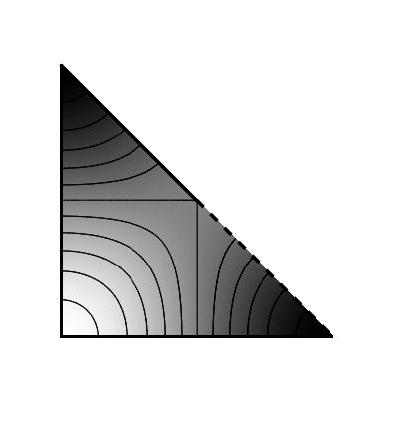\vspace*{0pt}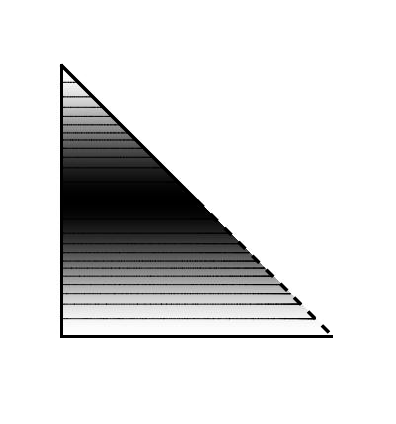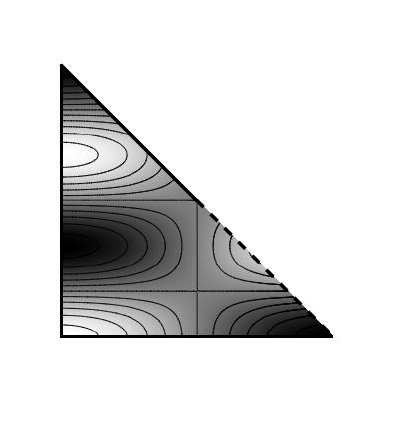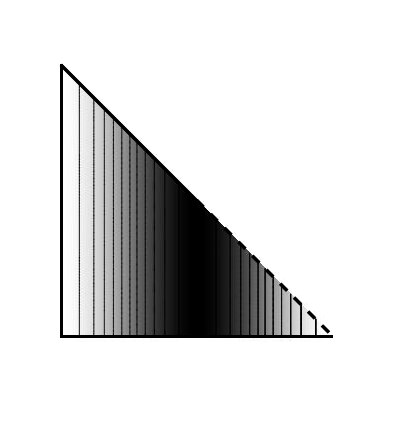
\caption{The contour plots of $\Xi^{l+}-$functions of $C_2$ over the fundamental domain $F^{l+}(C_2)$. The dashed part of the boundary does not belong to the fundamental domain.}\label{ElC2}
\end{figure}

For a point with coordinates in $\alpha^\vee$-basis $(x,y)$ we have the following explicit form of $\Xi^{l+}-$functions of $C_2$:
$$
\Xi^{l+}_{(a,b)}(x,y)=2\left\lbrace\cos(2\pi(ax+by))+\cos(2\pi(ax-(a+b)y))\right\rbrace.
$$
The fundamental domain $F^{l+}$ is of the form
$$F^{l+}(C_2) =\left\lbrace x\o_s^\vee+y\o_l^\vee\, |\, x,\, y\ge 0,\, 2x+y\le 1\right\rbrace\cup \left\lbrace (x+y)\o_s^\vee-y\o_l^\vee\, |\,x\ge 0,\,y> 0,\,2x+y< 1 \right\rbrace
$$ 
and the lattice of weights $P_{l+}$ is given by
$$P_{l+}(C_2) =\left\lbrace a\o_s+b\o_l\, |\,a,\, b\in \Z^{\ge0}\right\rbrace\cup \left\lbrace(a+2b)\o_s-b\o_l\, |\, a\in \Z^{\ge0},\, b\in \N\right\rbrace .
$$ 
The contour plots of some lowest $\Xi^{l+}-$functions of $C_2$ are given in Figure \ref{ElC2}.
The coefficients $d_\la^l$ of continuous orthogonality relations (\ref{l+corthog}) are given in Table \ref{tabStab}.

The discrete grid $F_M^{l+}$ is given by
\begin{align*}F_M^{l+}(C_2) =&\left\lbrace\dfrac{a}{M}\o_s^\vee+\dfrac{b}{M}\o_l^\vee \, |\,a,\,b,\,c\in \Z^{\ge0},\, c+2a+b= M\right\rbrace\\ 
&\cup \left\lbrace\dfrac{a+b}{M}\o_s^\vee-\dfrac{b}{M}\o_l^\vee \, |\,a\in \Z^{\ge0},\, b,\,c\in \N,\,c+ 2a+b=M\right\rbrace
\end{align*}
and the corresponding finite set of weights has the form
\begin{align*}\Lambda_M^{l+}(C_2) =&\left\lbrace a\o_s+b\o_l\, |\,a,\,b,\,c\in \Z^{\ge0},\, c+a+2b= M\right\rbrace\\ 
&\cup \left\lbrace(a+2b)\o_s-b\o_l\, |\, a,\,c\in \Z^{\ge0},\, b\in \N,\,c+ a+2b= M\right\rbrace.
\end{align*}
The coefficients $\ep^{l}(x)$ and $h_\la^{l\vee }$ of discrete orthogonality relations (\ref{l+dortho}) are given in Table \ref{tabOrb}. 

\subsubsection{$\Xi^{l+}-$functions of $G_2$}
\begin{figure}
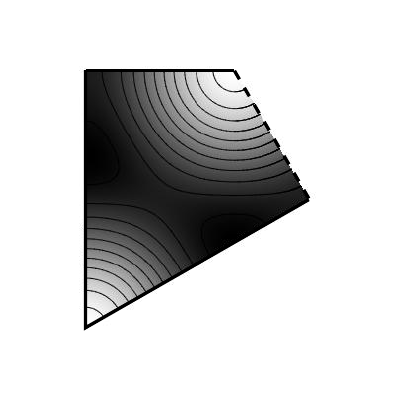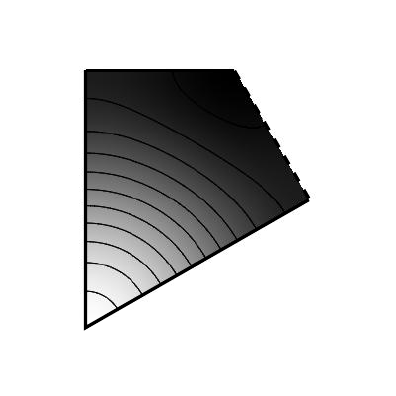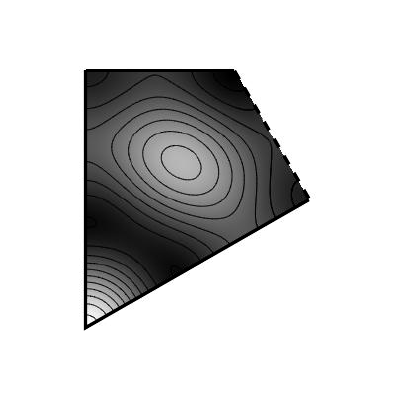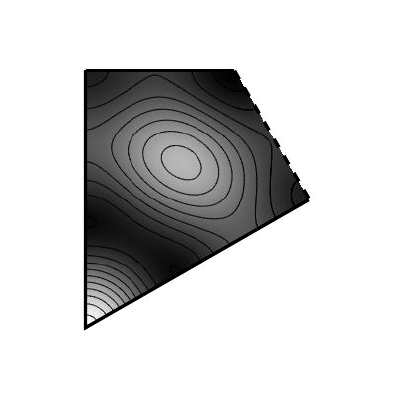\\  \vspace{-20pt}
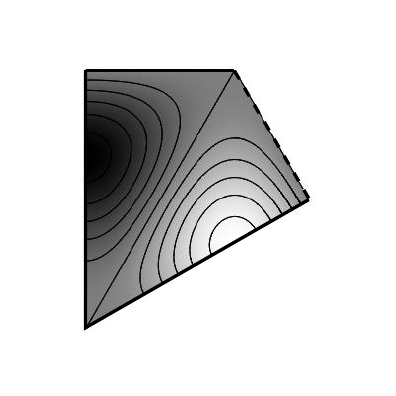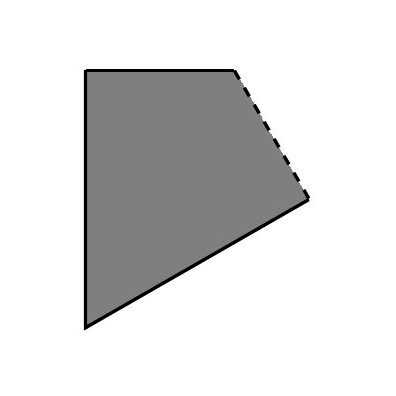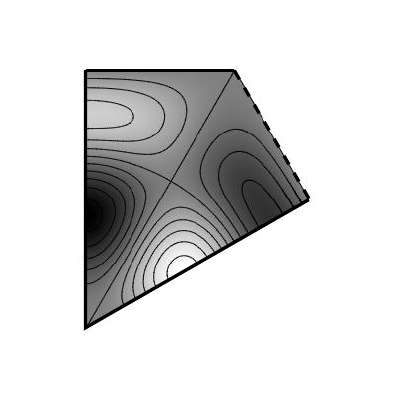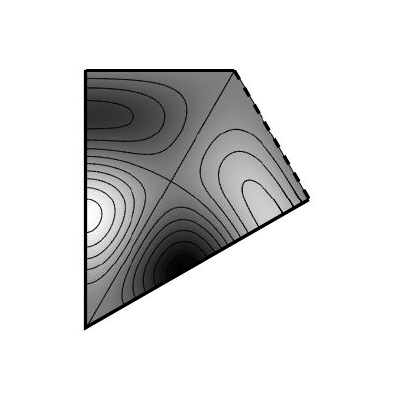
\caption{The contour plots of $\Xi^{l+}-$functions of $G_2$ over the fundamental domain $F^{l+}(G_2)$. The dashed part of the boundary does not belong to the fundamental domain.}\label{ElG2}
\end{figure}

For a point with coordinates in $\alpha^\vee$-basis $(x,y)$ we have the following explicit form of $\Xi^{l+}-$functions of $G_2$:
\begin{align*}
\Xi^{l+}_{(a,b)}(x,y)=&e^{2\pi i(ax+by)}+ e^{2\pi i ((a+b)x-by)}+ e^{2\pi i (-(2a+b)x+(3a+2b)y)} \\&+ e^{2\pi i ((a+b)x-(3a+2b)y)}+ e^{2\pi i (-(2a+b)x+(3a+b)y)}+ e^{2\pi i (ax-(3a+b)y)}.
\end{align*}
The fundamental domain $F^{l+}$ is of the form
$$F^{l+}(G_2)
=\left\lbrace x\o_l^\vee+y\o_s^\vee \, |\,x,\,y\ge0,\,2x+3y\le1\right\rbrace\cup \left\lbrace -x\o_l^\vee+(x+y)\o_s^\vee\, |\,x>0,\,y\ge0,\,2x+3y<1\right\rbrace
$$ 
and the lattice of weights $P_{l+}$ is given by
$$P_{l+}(G_2) =\left\lbrace  a\o_l+b\o_s\, |\,a,\,b\in\Z^{\ge0}\right\rbrace\cup \left\lbrace -a\o_l+(3a+b)\o_s \, |\,a\in\N,\,b\in\Z^{\ge0}\right\rbrace.
$$ 
The contour plots of some lowest $\Xi^{l+}-$functions of $G_2$ are given in Figure \ref{ElG2}.
The coefficients $d_\la^l$ of continuous orthogonality relations (\ref{l+corthog}) are given in Table \ref{tabStab}.

The discrete grid $F_M^{l+}$ is given by
\begin{align*}F_M^{l+}(G_2) =&\left\lbrace \dfrac{a}{M}\o_l^\vee+\dfrac{b}{M}\o_s^\vee\, |\,a,\,b,\,c\in \Z^{\ge0},\, c+2a+3b= M,\right\rbrace\\ 
&\cup \left\lbrace-\dfrac{a}{M}\o_l^\vee+\dfrac{a+b}{M}\o_s^\vee \, |\,a,\,c\in \N,\, b\in \Z^{\ge0},\,c+ 2a+3b= M\right\rbrace
\end{align*}
and the corresponding finite set of weights has the form
\begin{align*}\Lambda_M^{l+}(G_2) =&\left\lbrace a\o_l+b\o_s\, |\,a,\,b,\,c\in \Z^{\ge0}, c+3a+2b= M\right\rbrace\\ 
&\cup \left\lbrace -a\o_l+(3a+b)\o_s \, |\,a\in \N,\, b,\,c\in \Z^{\ge0},\, c+3a+2b= M\right\rbrace.
\end{align*}
The coefficients $\ep^{l}(x)$ and $h_\la^{l\vee }$ of discrete orthogonality relations (\ref{l+dortho}) are given in Table \ref{tabOrb}. 

\section{Mixed even orbit functions}\

Considering two different homomorphisms $\sigma,\wt\sigma\neq \id$  and corresponding kernels $\ker \sigma$, $\ker \wt\sigma$, we may define another type of 'mixed even' orbit functions $\varPsi^{\sigma, \wt\sigma}_\la:\R^2\map \C$ parametrized by $\la\in P$, $\sigma\neq \wt \sigma$ 
\begin{equation}\label{genorbevenmix}
\varPsi^{\sigma, \wt\sigma}_\la(x)=\sum_{w\in \ker \sigma }\wt\sigma (w) e^{2 \pi i \sca{ w\la}{x}}.
\end{equation} 
These 'mixed even' orbit functions are in general invariant or anti-invariant with respect to the action of $w\in\ker \sigma$ 
\begin{align}
\varPsi^{\sigma, \wt\sigma}_\la(wx)=\wt\sigma (w)\varPsi^{\sigma, \wt\sigma}_\la (x) \label{geninvevenmix1}\\
\varPsi^{\sigma, \wt\sigma}_{w\la}(x)=\wt\sigma (w)\varPsi^{\sigma, \wt\sigma}_\la (x)\label{geninvevenmix2}
\end{align}
and the invariance with respect to shifts from $Q^\vee$ also holds
\begin{equation}\label{genshiftevenmix}
\varPsi^{\sigma, \wt\sigma}_\la(x+q^\vee )=\varPsi^{\sigma, \wt\sigma}_\la (x),\q q^\vee \in Q^\vee.
\end{equation}
We have the following equalities of the mixed even orbit functions which correspond to all possible choices of the sign homomorphisms 
$$ \varPsi^{\sigma^e, \sigma^s}_\la = \varPsi^{\sigma^e, \sigma^l}_\la,\q \varPsi^{\sigma^l, \sigma^e}_\la = \varPsi^{\sigma^l, \sigma^s}_\la,\q \varPsi^{\sigma^s, \sigma^e}_\la = \varPsi^{\sigma^s, \sigma^l}_\la. $$
Thus, the mixed even orbit functions naturally split into three distinct classes.

\subsection{$\Xi^{e-}-$functions}\

We discuss in detail the functions $\varPsi^{\sigma^e, \sigma^s}_\la = \varPsi^{\sigma^e, \sigma^l}_\la$; 
we denote these functions by $\Xi_\la^{e-},\, \la \in P$ and the corresponding kernel is $W^e$. The (anti)invariance (\ref{geninvevenmix1}), (\ref{genshiftevenmix}) implies that these functions have common zeros in $F$:
\begin{equation}\label{zeroe}
\Xi_\la^{e-}(x)=0,\q x\in (Y_l \cup Y_0)\cap Y_s.
\end{equation}
Taking into account (\ref{geninvevenmix1}), (\ref{genshiftevenmix}) together with (\ref{zeroe}), we restrict the functions
$\Xi^{e-}_\la$ to the domain
\begin{equation*}
F^{e-}= ( F\setminus\{(Y_l \cup Y_0)\cap Y_s\})\, \cup\, r_s F^\circ.
\end{equation*} 
Similarly, the invariance (\ref{geninvevenmix2}) restricts $\la\in P$ to the set
\begin{equation*}
P_{e-}= (P^{+}\setminus (P^l\cap P^s)) \cup r_s P^{++}.
\end{equation*} 
Thus, we have
\begin{equation*}
\Xi^{e-}_\la(x)=\sum_{w\in W^e } \sigma^s(w) e^{2 \pi i \sca{ w\la}{x}},\q x\in F^{e-},\, \la \in P_{e-}.
\end{equation*} 

\subsubsection{Continuous orthogonality and $\Xi^{e-}-$transforms}
For any two weights $\la,\la'\in P_{e-}$ are the
corresponding $\Xi^{e-}-$functions orthogonal on $F^{e-}$
\begin{equation}\label{e-corthog}
\int_{{F}^{e-}}\Xi^{e-}_{\lambda}(x)\overline{\Xi^{e-}_{\lambda'}(x)}\,dx=
K\,d^{e}_{\la}\,\delta_{\lambda\lambda'} \end{equation}
where $d^{e}_{\la}$, $K$ are given by (\ref{dla}), (\ref{K}), respectively. The $\Xi^{e-}-$functions determine symmetrized Fourier series
expansions,
\begin{equation*}
f(x)=\sum_{\la\in P_{e-}}c^{e-}_{\la}\Xi^{e-}_{\la}(x),\quad {\mathrm{
where}}\
c^{e-}_{\la}=\frac{1}{K d^{e}_{\la}}\int_{F^{e-}}f(x)\overline{\Xi^{e-}_{\la}(x)}\,dx.
\end{equation*}

\subsubsection{Discrete orthogonality and discrete $\Xi^{e-}-$transforms}\ The finite set of points is given by $$F_M^{e-}=\frac{1}{M}P^\vee / Q^\vee \cap F^{e-}.  $$
We define the corresponding finite set of weights as 
$$\Lambda^{e-}_M = P/MQ \cap MF^{e-\vee}$$
where 
$$F^{e-\vee}=(F^\vee\setminus \{ (Y_s^\vee\cup Y_0^\vee)\cap Y_l^\vee \})\, \cup \,r_s F^{\vee\circ}.$$
Then, for $\la,\la' \in\Lambda^{e-}_M$ ,the following discrete orthogonality relations hold
\begin{equation}\label{e-dortho}
 \sum_{x\in F^{e-}_M}\ep^e(x) \Xi^{e-}_\la(x)\overline{\Xi^{e-}_{\la'}(x)}=kM^2 h^{e\vee}_\la \delta_{\la\la'}
\end{equation}
where $h^{e\vee}_{\la}$, $k$ are given by (\ref{hla}), (\ref{k}), respectively. The discrete symmetrized $\Xi^{e-}-$functions expansion is given by                    \begin{equation*}
f(x)=\sum_{\la\in \Lambda_M^{e-}}c^{e-}_{\la}\Xi^{e-}_{\la}(x),\quad {\mathrm{
where}}\
c^{e-}_{\la}=\frac{1}{k M^2 h^{e\vee}_\la}\sum_{x\in F^{e-}_M}\ep^{e}(x) f(x)\overline{\Xi^{e-}_{\la}(x)}.
\end{equation*}

\subsubsection{$\Xi^{e-}-$functions of $C_2$}
\begin{figure}
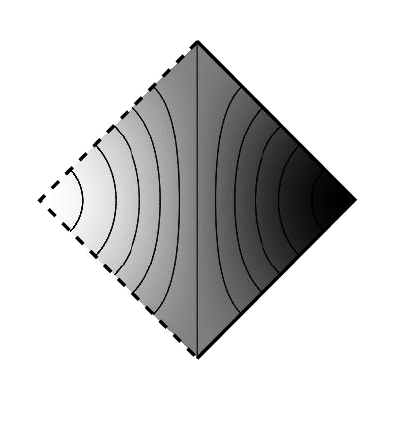\vspace*{0pt}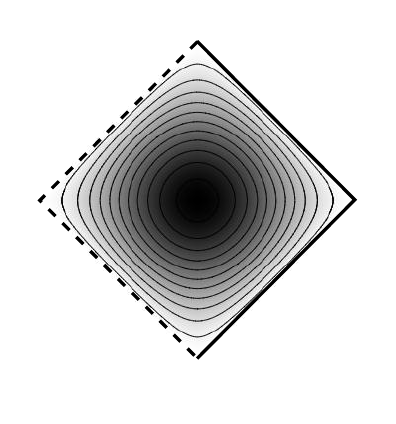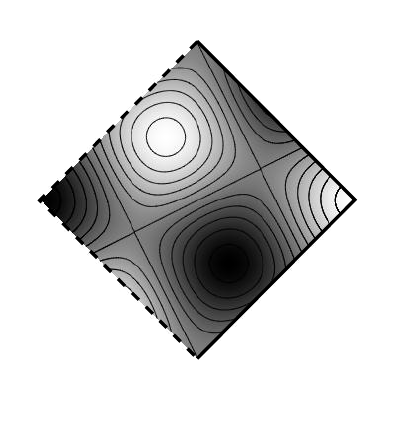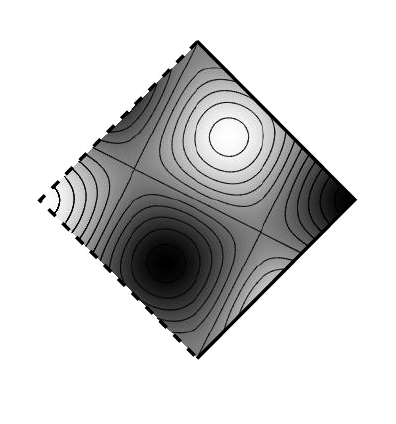
\caption{The contour plots of $\Xi^{e-}-$functions of $C_2$ over the fundamental domain $F^{e-}(C_2)$. The dashed part of the boundary does not belong to the fundamental domain.}\label{EemC2}
\end{figure}

For a point with coordinates in $\alpha^\vee$-basis $(x,y)$ we have the following explicit form of $\Xi^{e-}-$functions of $C_2$:
$$
\Xi^{e-}_{(a,b)}(x,y)=2\left\lbrace\cos(2\pi(ax+by))-\cos(2\pi((a+2b)x-(a+b)y))\right\rbrace.
$$
The fundamental domain $F^{e-}$ is of the form
\begin{align*}F^{e-}(C_2) =&\left\lbrace x\o_s^\vee+y\o_l^\vee\, |\, x,\, y\ge 0, 2x+y\le 1,\, (x,y)\neq (0,0),(0,1)\right\rbrace \\
&\cup \left\lbrace -x\o_s^\vee+(2x+y)\o_l^\vee\, |\,x,\,y> 0, 2x+y< 1 \right\rbrace
\end{align*}
and the lattice of weights $P_{e-}$ is given by
$$P_{e-}(C_2) =\left\lbrace a\o_s+b\o_l\, |\,a,\, b\in \Z^{\ge0},\, (a,b)\neq (0,0)\right\rbrace 
\cup \left\lbrace-a\o_s+(a+b)\o_l\, |\, a,\, b\in \N\right\rbrace .
$$ 
The contour plots of some lowest $\Xi^{e-}-$functions of $C_2$ are given in Figure \ref{EemC2}.
The coefficients $d_\la^e$ of continuous orthogonality relations (\ref{e-corthog}) are given in Table \ref{tabStab}.

The discrete grid $F_M^{e-}$ is given by
\begin{align*}F_M^{e-}(C_2) =&\left\lbrace\dfrac{a}{M}\o_s^\vee+\dfrac{b}{M}\o_l^\vee \, |\,c,\,a,\, b\in \Z^{\ge0},\, c+2a+b= M,\, [c,a,b]\neq [M,0,0],[0,0,M]\right\rbrace\\ 
&\cup \left\lbrace-\dfrac{a}{M}\o_s^\vee+\dfrac{2a+b}{M}\o_l^\vee \, |\,a,\,b,\,c\in \N,\, c+2a+b= M\right\rbrace
\end{align*}
and the corresponding finite set of weights has the form
\begin{align*}\Lambda_M^{e-}(C_2) =&\left\lbrace a\o_s+b\o_l\, |\, c,\,a,\,b \in \Z^{\ge0},\, c+a+2b= M,\, [c,a,b]\neq [M,0,0],[0,M,0]\right\rbrace\\ 
&\cup \left\lbrace-a\o_s+(a+b)\o_l\, |\, a,\,b,\,c\in \N,\,c+ a+2b= M\right\rbrace.
\end{align*}
The coefficients $\ep^{e}(x)$ and $h_\la^{e\vee }$ of discrete orthogonality relations (\ref{e-dortho}) are given in Table \ref{tabOrb}. 

\subsubsection{$\Xi^{e-}-$functions of $G_2$}
\begin{figure}
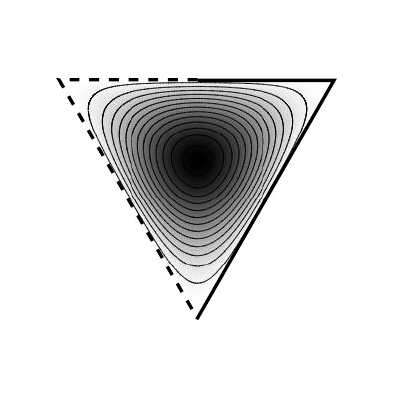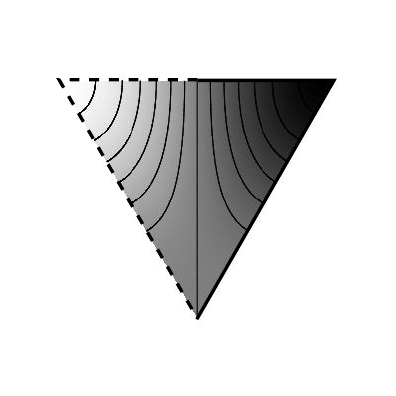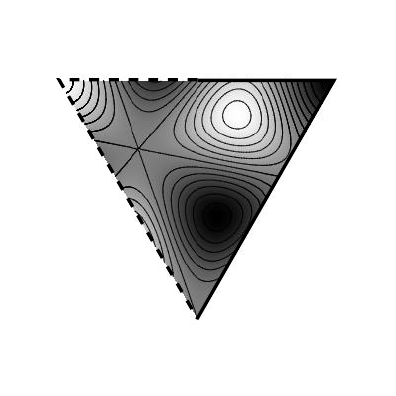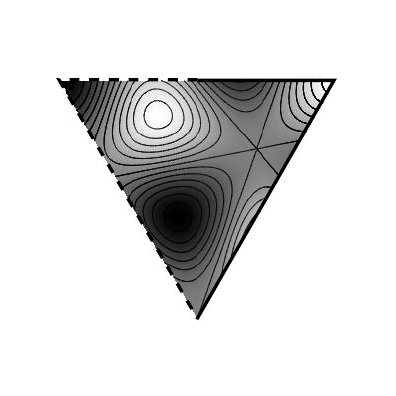\\ 
\caption{The contour plots of $\Xi^{e-}-$functions of $G_2$ over the fundamental domain $F^{e-}(G_2)$. The dashed part of the boundary does not belong to the fundamental domain. Real parts of $\Xi^{e-}$-functions are zero.}\label{EmG2}
\end{figure}

For a point with coordinates in $\alpha^\vee$-basis $(x,y)$ we have the following explicit form of $\Xi^{e-}-$functions of $G_2$:
\begin{align*}
\Xi^{e-}_{(a,b)}(x,y)=&2i\lbrace\sin(2\pi(ax+by))+\sin(2\pi((3a+b)y)-(2a+b)x)\\ 
& +\sin(2\pi((a+b)x-(3a+2b)x))\rbrace.
\end{align*}
The fundamental domain $F^{e-}$ is of the form
\begin{align*}F^{e-}(G_2)
=&\left\lbrace x\o_l^\vee+y\o_s^\vee \, |\,x,\,y\ge0,2x+3y\le 1,\, (x,y)\neq (0,0),(1/2,0)\right\rbrace \\ & \cup \left\lbrace (x+3y)\o_l^\vee-y\o_s^\vee\, |\,x,\,y>0,\,2x+3y<1\right\rbrace
\end{align*}
and the lattice of weights $P_{e-}$ is given by
$$P_{e-}(G_2) =\left\lbrace  a\o_l+b\o_s\, |\,a,\,b\in\Z^{\ge0},\,(a,b)\neq (0,0)\right\rbrace
\cup \left\lbrace (a+b)\o_l-b\o_s \, |\,a,\,b\in\N\right\rbrace
$$ 
The contour plots of some lowest $\Xi^{e-}-$functions of $G_2$ are given in Figure \ref{EmG2}.

The coefficients $d_\la^e$ of continuous orthogonality relations (\ref{e-corthog}) are given in Table \ref{tabStab}.
The discrete grid $F_M^{e-}$ is given by
\begin{align*}F_M^{e-}(G_2) =&\left\lbrace \dfrac{a}{M}\o_l^\vee+\dfrac{b}{M}\o_s^\vee \,\mid\, a,\,b,\,c\in \Z^{\ge0},\, c+ 2a+3b= M,\, (c,a,b)\neq (M,0,0),\, (0,M/2,0) \right\rbrace\\ 
&\cup \left\lbrace\dfrac{a+3b}{M}\o_l^\vee-\dfrac{b}{M}\o_s^\vee \, |\,c,\,a,\, b\in \N,\, c+2a+3b= M\right\rbrace
\end{align*}
and the corresponding finite set of weights has the form
\begin{align*}\Lambda_M^{e-}(G_2) =&\left\lbrace a\o_l+b\o_s\, |\,a, \, b,\, c\in \Z^{\ge0},\,c+ 3a+2b= M,\, [c,a,b]\neq [M,0,0],\, [0,0,M/2] \right\rbrace\\ 
&\cup \left\lbrace (a+b)\o_l-b\o_s \, |\, a,\, b,\,c\in \N,\,c+ 3a+2b= M\right\rbrace.
\end{align*}
The coefficients $\ep^{e}(x)$ and $h_\la^{e\vee }$ of discrete orthogonality relations (\ref{e-dortho}) are given in Table \ref{tabOrb}. 

\subsection{$\Xi^{s-}-$functions}\

We discuss in detail the functions $\varPsi^{\sigma^s, \sigma^e}_\la = \varPsi^{\sigma^s, \sigma^l}_\la$; 
we denote these functions by $\Xi_\la^{s-},\, \la \in P$ and the corresponding kernel is $W^s$. The (anti)invariance (\ref{geninvevenmix1}), (\ref{genshiftevenmix}) implies that these functions have common zeros in $F$:
\begin{equation}\label{zeros}
\Xi_\la^{s-}(x)=0,\q x\in Y_l \cup Y_0.
\end{equation}
Taking into account (\ref{geninvevenmix1}), (\ref{genshiftevenmix}) together with (\ref{zeros}), we restrict the functions
$\Xi^{s-}_\la$ to the domain
\begin{equation*}
F^{s-}= ( F\setminus (Y_l \cup Y_0))\, \cup\, r_s F^\circ.
\end{equation*} 
Similarly, the invariance (\ref{geninvevenmix2}) restricts $\la\in P$ to the set
\begin{equation*}
P_{s-}= (P^{+}\setminus P_l)\, \cup\, r_s P^{++}.
\end{equation*} 
Thus, we have
\begin{equation*}
\Xi^{s-}_\la(x)=\sum_{w\in W^s }\sigma^l (w) e^{2 \pi i \sca{ w\la}{x}},\q x\in F^{s-},\, \la \in P_{s-}.
\end{equation*} 

\subsubsection{Continuous orthogonality and $\Xi^{s-}-$transforms}
For any two weights $\la,\la'\in P_{s-}$ the
corresponding $\Xi^{s-}-$functions are orthogonal on $F^{s-}$
\begin{equation}\label{s-corthog}
\int_{{F}^{s-}}\Xi^{s-}_{\lambda}(x)\overline{\Xi^{s-}_{\lambda'}(x)}\,dx=
K\,\delta_{\lambda\lambda'} \end{equation}
where $K$ is given by (\ref{K}). The $\Xi^{s-}-$functions determine symmetrized Fourier series
expansions,
\begin{equation*}
f(x)=\sum_{\la\in P_{s-}}c^{s-}_{\la}\Xi^{s-}_{\la}(x),\quad {\mathrm{
where}}\
c^{s-}_{\la}=\frac{1}{K}\int_{F^{s-}}f(x)\overline{\Xi^{s-}_{\la}(x)}\,dx.
\end{equation*}

\subsubsection{Discrete orthogonality and discrete $\Xi^{s-}-$transforms}\ The finite set of points is given by $$F_M^{s-}=\frac{1}{M}P^\vee / Q^\vee \cap F^{s-}.  $$
We define the corresponding finite set of weights as $$\Lambda^{s-}_M = P/MQ \cap MF^{s-\vee} $$
where $$F^{s-\vee}=(F^\vee \setminus Y_l^\vee)\, \cup\, r_s F^{\vee \circ}.$$
Then, for $\la,\la' \in\Lambda^{s-}_M$, the following discrete orthogonality relations hold
\begin{equation}\label{s-dortho}
 \sum_{x\in F^{s-}_M}\ep^s(x) \Xi^{s-}_\la(x)\overline{\Xi^{s-}_{\la'}(x)}=kM^2 h^{s\vee}_\la \delta_{\la\la'}
\end{equation}
where $h^{s\vee}_{\la}$, $k$ are given by (\ref{hla}), (\ref{k}), respectively. The discrete symmetrized $\Xi^{s-}-$functions expansion is given by                    \begin{equation*}
f(x)=\sum_{\la\in \Lambda_M^{s-}}c^{s-}_{\la}\Xi^{s-}_{\la}(x),\quad {\mathrm{
where}}\
c^{s-}_{\la}=\frac{1}{k M^2 h^{s\vee}_\la}\sum_{x\in F^{s-}_M}\ep^{s}(x) f(x)\overline{\Xi^{s-}_{\la}(x)}.
\end{equation*}

\subsubsection{$\Xi^{s-}-$functions of $C_2$}
\begin{figure}
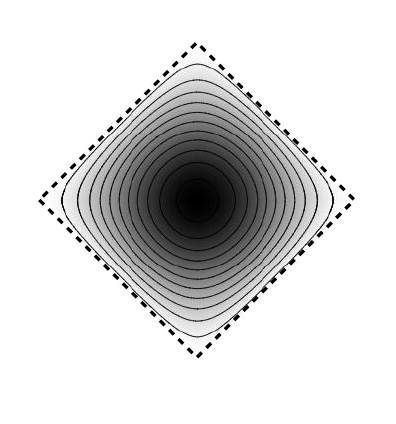\vspace*{0pt}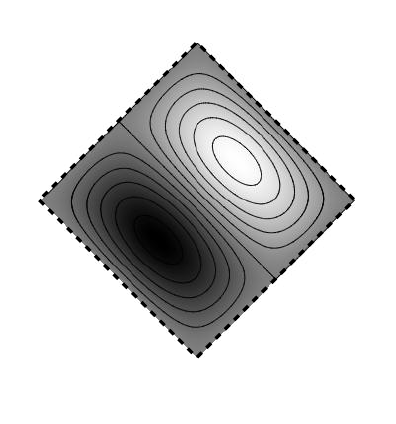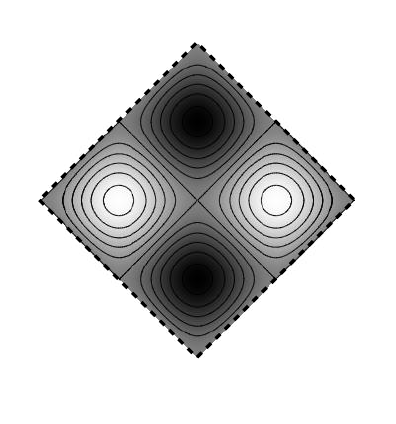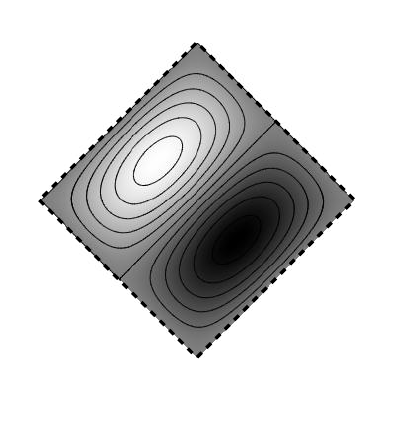
\caption{The contour plots of $\Xi^{s-}-$functions of $C_2$ over the fundamental domain $F^{s-}(C_2)$. The dashed part of the boundary does not belong to the fundamental domain.}\label{EsmC2}
\end{figure}

For a point with coordinates in $\alpha^\vee$-basis $(x,y)$ we have the following explicit form of $\Xi^{s-}-$functions of $C_2$:
$$
\Xi^{s-}_{(a,b)}(x,y)=2\left\lbrace\cos(2\pi(ax+by))-\cos(2\pi((a+2b)x-by))\right\rbrace.
$$
The fundamental domain $F^{s-}$ is of the form
$$F^{s-}(C_2) =\left\lbrace x\o_s^\vee+y\o_l^\vee\, |\, x\ge 0,\, y>0,\, 2x+y < 1\right\rbrace 
\cup \left\lbrace -x\o_s^\vee+(2x+y)\o_l^\vee\, |\,x,\,y > 0,\,  2x+y< 1 \right\rbrace
$$ 
and the lattice of weights $P_{s-}$ is given by
$$P_{s-}(C_2) =\left\lbrace a\o_s+b\o_l\, |\,a\in \Z^{\ge0},\, b\in \N \right\rbrace 
\cup \left\lbrace-a\o_s+(a+b)\o_l\, |\, a,\, b\in \N\right\rbrace .
$$ 
The contour plots of some lowest $\Xi^{s-}-$functions of $C_2$ are given in Figure \ref{EsmC2}.

The discrete grid $F_M^{s-}$ is given by
\begin{align*}F_M^{s-}(C_2) =&\left\lbrace\dfrac{a}{M}\o_s^\vee+\dfrac{b}{M}\o_l^\vee \, |\,a\in \Z^{\ge0},\,b,\,c\,\in \N,\, c+2a+b= M\right\rbrace\\ 
&\cup \left\lbrace-\dfrac{a}{M}\o_s^\vee+\dfrac{2a+b}{M}\o_l^\vee \, |\,a,\,b,\,c\in \N,\, c+2a+b= M\right\rbrace
\end{align*}
and the corresponding finite set of weights has the form
\begin{align*}\Lambda_M^{s-}(C_2) =&\left\lbrace a\o_s+b\o_l\, |\, c,\,a \in \Z^{\ge0},\,b\in\N,\, c+a+2b= M\right\rbrace\\ 
&\cup \left\lbrace-a\o_s+(a+b)\o_l\, |\, a,\,b,\,c\in \N,\,c+a+2b= M\right\rbrace.
\end{align*}
The coefficients $\ep^{s}(x)$ of discrete orthogonality relations (\ref{s-dortho}) have a common value $\ep^{s}(x)=4$ for all $x\in F_M^{s-}(C_2)$. The coefficients $h_\la^{s\vee }$ are given in Table \ref{tabOrb}. 

\subsubsection{$\Xi^{s-}-$functions of $G_2$}
\begin{figure}
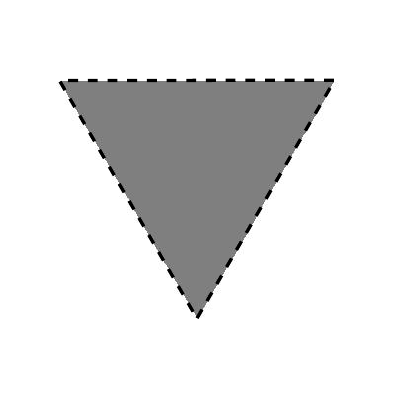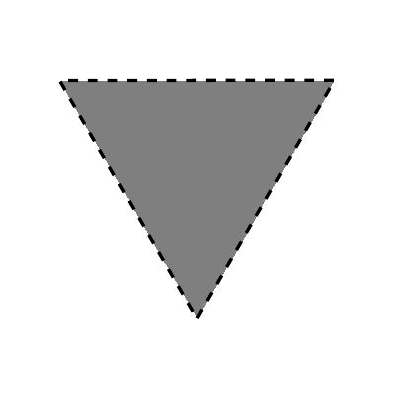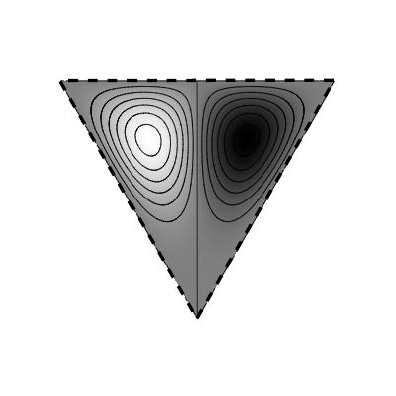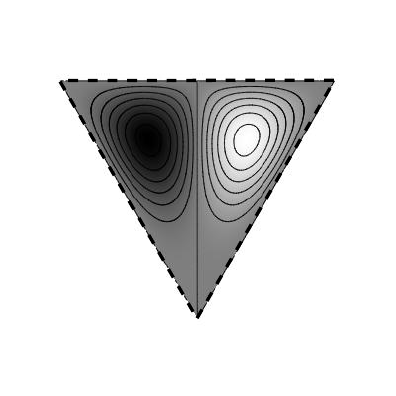\\ \vspace{-20pt}
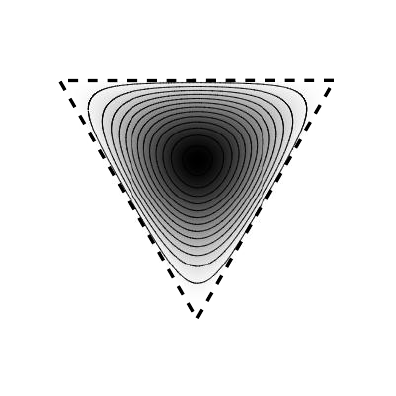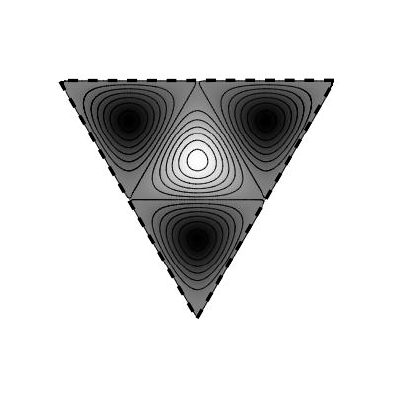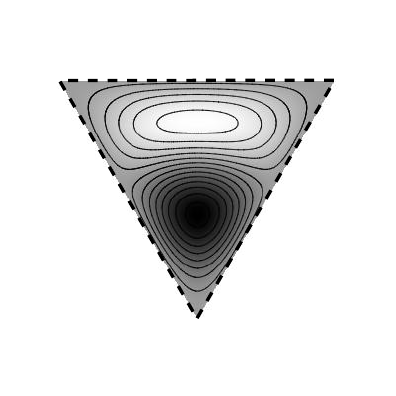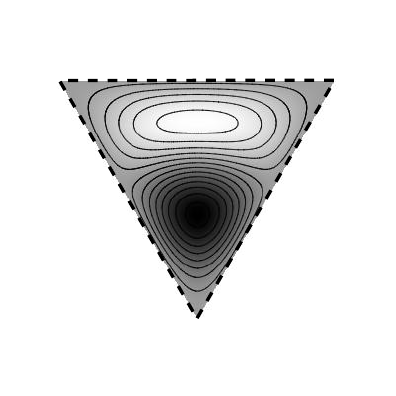
\caption{The contour plots of $\Xi^{s-}-$functions of $G_2$ over the fundamental domain $F^{s-}(G_2)$. The dashed part of the boundary does not belong to the fundamental domain.}\label{EsmG2}
\end{figure}

For a point with coordinates in $\alpha^\vee$-basis $(x,y)$ we have the following explicit form of $\Xi^{s-}-$functions of $G_2$:
\begin{align*}
\Xi^{s-}_{(a,b)}(x,y)=&e^{2\pi i(ax+by)}- e^{2\pi i (-ax+(3a+b)y)}- e^{2\pi i ((2a+b)x-(3a+2b)y)}\\ &+ e^{2\pi i ((a+b)x-(3a+2b)y)}+ e^{2\pi i (-(2a+b)x+(3a+b)y)}- e^{2\pi i (-(a+b)x+by)}.
\end{align*}
The fundamental domain $F^{s-}$ is of the form
$$F^{s-}(G_2)
=\left\lbrace x\o_l^\vee+y\o_s^\vee \, |\,x>0,\,y\ge0,2x+3y<1\right\rbrace\cup \left\lbrace (x+3y)\o_l^\vee-y\o_s^\vee\, |\,x,\,y>0,\,2x+3y<1\right\rbrace\\
$$ 
and the lattice of weights $P_{s-}$ is given by
$$P_{s-}(G_2) =\left\lbrace  a\o_l+b\o_s\, |\,a\in\N,\,b\in\Z^{\ge0}\right\rbrace
\cup \left\lbrace (a+b)\o_l-b\o_s \, |\,a,\,b\in\N\right\rbrace.
$$ 
The contour plots of some lowest $\Xi^{s-}-$functions of $G_2$ are given in Figure \ref{EsmG2}.

The discrete grid $F_M^{s-}$ is given by
\begin{align*}F_M^{s-}(G_2) =&\left\lbrace \dfrac{a}{M}\o_l^\vee+\dfrac{b}{M}\o_s^\vee \,\mid\,b\in \Z^{\ge0},\,a,\,c\in\N,\, c+ 2a+3b= M\right\rbrace\\ 
&\cup \left\lbrace\dfrac{a+3b}{M}\o_l^\vee-\dfrac{b}{M}\o_s^\vee \, |\,a,\, b,\,c\in \N,\, c+2a+3b= M\right\rbrace
\end{align*}
and the corresponding finite set of weights has the form
\begin{align*}\Lambda_M^{s-}(G_2) =&\left\lbrace a\o_l+b\o_s\, |\,a\in \N, \, b,\, c\in \Z^{\ge0},\,c+ 3a+2b= M\right\rbrace\\ 
&\cup \left\lbrace (a+b)\o_l-b\o_s \, |\, a,\, b,\,c\in \N,\,c+ 3a+2b= M\right\rbrace.
\end{align*}
The coefficients $\ep^{s}(x)$ of discrete orthogonality relations (\ref{s-dortho}) have a common value $\ep^{s}(x)=6$ for all $x\in F_M^{s-}(G_2)$. The coefficients $h_\la^{s\vee }$ are given in Table \ref{tabOrb}. 

\subsection{$\Xi^{l-}-$functions}\

We discuss in detail the functions $\varPsi^{\sigma^l, \sigma^e}_\la = \varPsi^{\sigma^l, \sigma^s}_\la$; 
we denote these functions by $\Xi_\la^{l-},\, \la \in P$ and the corresponding kernel is $W^l$. The (anti)invariance (\ref{geninvevenmix1}), (\ref{genshiftevenmix}) implies that these functions have common zeros in $F$:
\begin{equation}\label{zerol}
\Xi_\la^{l-}(x)=0,\q x\in Y_s.
\end{equation}
Taking into account (\ref{geninvevenmix1}), (\ref{genshiftevenmix}) together with (\ref{zerol}), we restrict the functions
$\Xi^{l-}_\la$ to the domain
\begin{equation*}
F^{l-}= ( F\setminus Y_s)\, \cup\, r_l F^\circ.
\end{equation*} 
Similarly, the invariance (\ref{geninvevenmix2}) restricts $\la\in P$ to the set
\begin{equation*}
P_{l-}= (P^{+}\setminus P_s)\, \cup\, r_l P^{++}.
\end{equation*} 
Thus, we have
\begin{equation*}
\Xi^{l-}_\la(x)=\sum_{w\in W^l }\sigma^s (w) e^{2 \pi i \sca{ w\la}{x}},\q x\in F^{l-},\, \la \in P_{l-}.
\end{equation*} 

\subsubsection{Continuous orthogonality and $\Xi^{l-}-$transforms}
For any two weights $\la,\la'\in P_{l-}$ the
corresponding $\Xi^{l-}-$functions are orthogonal on $F^{l-}$
\begin{equation}\label{l-corthog}
\int_{{F}^{l-}}\Xi^{l-}_{\lambda}(x)\overline{\Xi^{l-}_{\lambda'}(x)}\,dx=
K\,\delta_{\lambda\lambda'} \end{equation}
where $K$ is given by (\ref{K}). The $\Xi^{l-}-$functions determine symmetrized Fourier series
expansions,
\begin{equation*}
f(x)=\sum_{\la\in P_{l-}}c^{l-}_{\la}\Xi^{l-}_{\la}(x),\quad {\mathrm{
where}}\
c^{l-}_{\la}=\frac{1}{K}\int_{F^{l-}}f(x)\overline{\Xi^{l-}_{\la}(x)}\,dx.
\end{equation*}

\subsubsection{Discrete orthogonality and discrete $\Xi^{l-}-$transforms}\ The finite set of points is given by $$F_M^{l-}=\frac{1}{M}P^\vee / Q^\vee \cap F^{l-}.  $$
We define the corresponding finite set of weights as $$\Lambda^{l-}_M = P/MQ \cap MF^{l-\vee} $$
where $$F^{l-\vee}=(F^\vee\setminus (Y^\vee_s\cup Y^\vee_0  ))\, \cup r_l F^{\vee\circ} .$$
Then, for $\la,\la' \in\Lambda^{l-}_M$, the following discrete orthogonality relations hold
\begin{equation}\label{l-dortho}
 \sum_{x\in F^{l-}_M}\ep^l(x) \Xi^{l-}_\la(x)\overline{\Xi^{l-}_{\la'}(x)}=k M^2 \delta_{\la\la'}
\end{equation}
where $k$ is given by (\ref{k}). The discrete symmetrized $\Xi^{l-}-$functions expansion is given by                    \begin{equation*}
f(x)=\sum_{\la\in \Lambda_M^{l-}}c^{l-}_{\la}\Xi^{l-}_{\la}(x),\quad {\mathrm{
where}}\
c^{l-}_{\la}=\frac{1}{k M^2}\sum_{x\in F^{l-}_M}\ep^{l}(x) f(x)\overline{\Xi^{l-}_{\la}(x)}.
\end{equation*}

\subsubsection{$\Xi^{l-}-$functions of $C_2$}
\begin{figure}
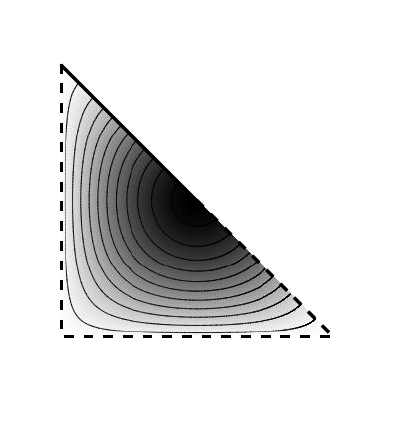\vspace*{0pt}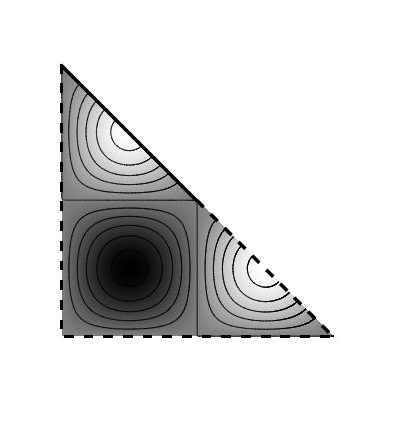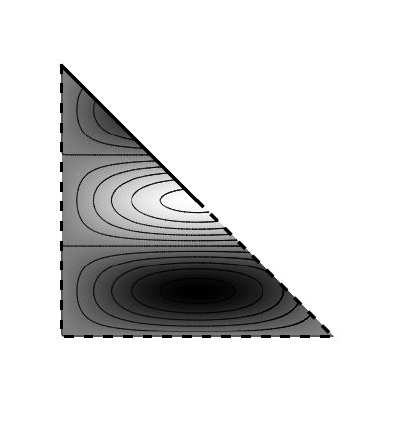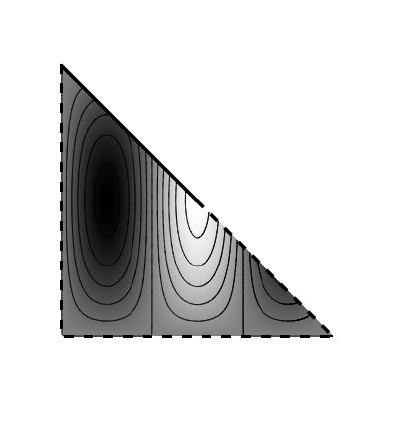
\caption{The contour plots of $\Xi^{l-}-$functions of $C_2$ over the fundamental domain $F^{l-}(C_2)$. The dashed part of the boundary does not belong to the fundamental domain.}\label{ElmC2}
\end{figure}

For a point with coordinates in $\alpha^\vee$-basis $(x,y)$ we have the following explicit form of $\Xi^{l-}-$functions of $C_2$:
$$
\Xi^{l-}_{(a,b)}(x,y)=2\left\lbrace\cos(2\pi(ax+by))-\cos(2\pi(ax-(a+b)y))\right\rbrace.
$$
The fundamental domain $F^{l-}$ is of the form
$$F^{l-}(C_2) =\left\lbrace x\o_s^\vee+y\o_l^\vee\, |\, x>0,\, y\ge 0,\, 2x+y\le 1\right\rbrace\cup \left\lbrace (x+y)\o_s^\vee-y\o_l^\vee\, |\,x,\,y> 0,\,2x+y< 1 \right\rbrace
$$ 
and the lattice of weights $P_{l-}$ is given by
$$P_{l-}(C_2) =\left\lbrace a\o_s+b\o_l\, |\,a\in \N,\, b\in \Z^{\ge0}\right\rbrace\cup \left\lbrace(a+2b)\o_s-b\o_l\, |\, a,\, b\in \N\right\rbrace .
$$ 
The contour plots of some lowest $\Xi^{l-}-$functions of $C_2$ are given in Figure \ref{ElmC2}.

The discrete grid $F_M^{l-}$ is given by
\begin{align*}F_M^{l-}(C_2) =&\left\lbrace\dfrac{a}{M}\o_s^\vee+\dfrac{b}{M}\o_l^\vee \, |\,a\in \N,\,b,\,c\in \Z^{\ge0},\, c+2a+b= M\right\rbrace\\ 
&\cup \left\lbrace\dfrac{a+b}{M}\o_s^\vee-\dfrac{b}{M}\o_l^\vee \, |\,a,\, b,\,c\in \N,\,c+ 2a+b=M\right\rbrace
\end{align*}
and the corresponding finite set of weights has the form
\begin{align*}\Lambda_M^{l-}(C_2) =&\left\lbrace a\o_s+b\o_l\, |\,a,\,c\in\N,\,b\in \Z^{\ge0},\, c+a+2b= M\right\rbrace\\ 
&\cup \left\lbrace(a+2b)\o_s-b\o_l\, |\, a,\,b,\,c\in \N,\,c+ a+2b= M\right\rbrace.
\end{align*}
The coefficients $\ep^{l}(x)$ of discrete orthogonality relations (\ref{l-dortho}) are given in Table \ref{tabOrb}. 

\subsubsection{$\Xi^{l-}-$functions of $G_2$}
\begin{figure}
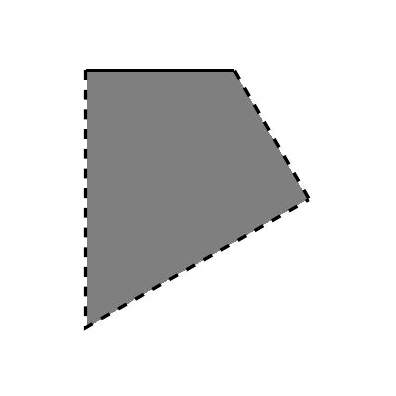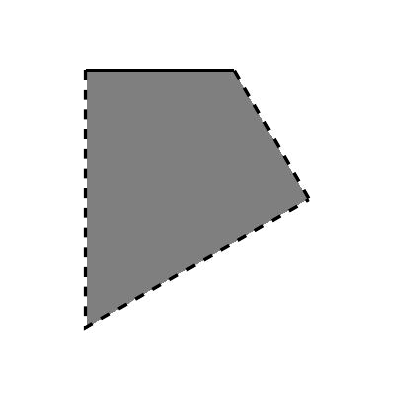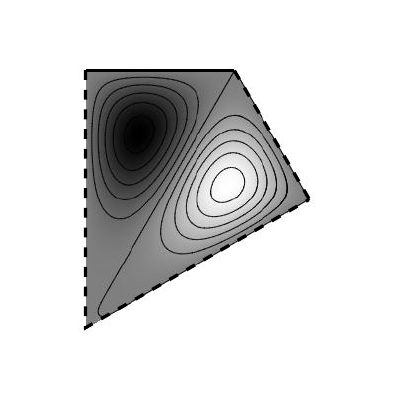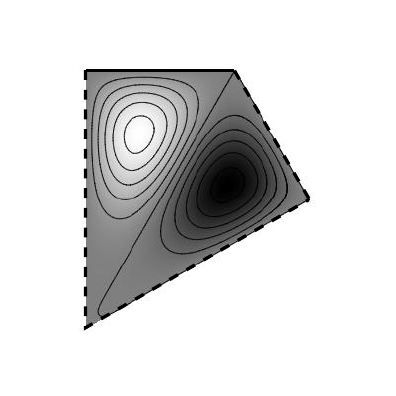\\  \vspace{-20pt}
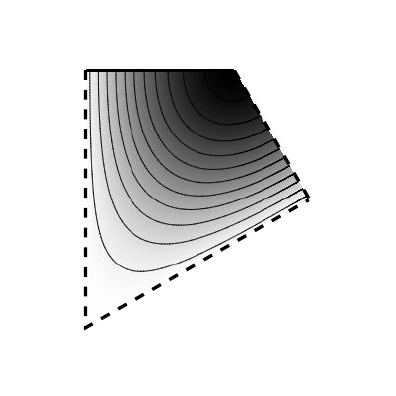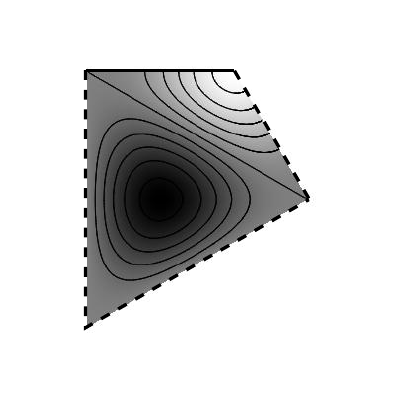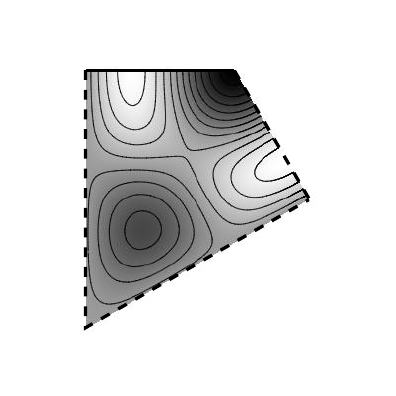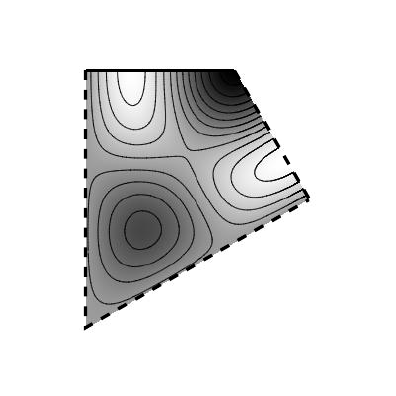
\caption{The contour plots of $\Xi^{l-}-$functions of $G_2$ over the fundamental domain $F^{l-}(G_2)$. The dashed part of the boundary does not belong to the fundamental domain.}\label{ElmG2}
\end{figure}

For a point with coordinates in $\alpha^\vee$-basis $(x,y)$ we have the following explicit form of $\Xi^{l-}-$functions of $G_2$:
\begin{align*}
\Xi^{l-}_{(a,b)}(x,y)=&e^{2\pi i(ax+by)}- e^{2\pi i ((a+b)x-by)}- e^{2\pi i (-(2a+b)x+(3a+2b)y)} \\&+ e^{2\pi i ((a+b)x-(3a+2b)y)}+ e^{2\pi i (-(2a+b)x+(3a+b)y)}- e^{2\pi i (ax-(3a+b)y)}.
\end{align*}
The fundamental domain $F^{l-}$ is of the form
$$F^{l-}(G_2)
=\left\lbrace x\o_l^\vee+y\o_s^\vee \, |\,x\ge0,\,y>0,\,2x+3y\le1\right\rbrace\cup \left\lbrace -x\o_l^\vee+(x+y)\o_s^\vee\, |\,x,\,y>0,\,2x+3y<1\right\rbrace
$$ 
and the lattice of weights $P_{l-}$ is given by
$$P_{l-}(G_2) =\left\lbrace  a\o_l+b\o_s\, |\,a\in\Z^{\ge0},\, b\in \N\right\rbrace\cup \left\lbrace -a\o_l+(3a+b)\o_s \, |\,a,\,b\in\N\right\rbrace.
$$ 
The contour plots of some lowest $\Xi^{l-}-$functions of $G_2$ are given in Figure \ref{ElmG2}.

The discrete grid $F_M^{l-}$ is given by
\begin{align*}F_M^{l-}(G_2) =&\left\lbrace \dfrac{a}{M}\o_l^\vee+\dfrac{b}{M}\o_s^\vee\, |\,a,\,c\in \Z^{\ge0},\,b\in\N,\, c+2a+3b= M,\right\rbrace\\ 
&\cup \left\lbrace-\dfrac{a}{M}\o_l^\vee+\dfrac{a+b}{M}\o_s^\vee \, |\,a,\,b,\,c\in \N,\,c+ 2a+3b= M\right\rbrace
\end{align*}
and the corresponding finite set of weights has the form
\begin{align*}\Lambda_M^{l-}(G_2) =&\left\lbrace a\o_l+b\o_s\, |\,a\in \Z^{\ge0},\,b,\,c\in \N,\, c+3a+2b= M\right\rbrace\\ 
&\cup \left\lbrace -a\o_l+(3a+b)\o_s \, |\,a,\, b,\,c\in \N,\, c+3a+2b= M\right\rbrace.
\end{align*}
The coefficients $\ep^{l}(x)$ of discrete orthogonality relations (\ref{l-dortho}) are given in Table \ref{tabOrb}. 

\section{Product decompositions}

Various products of generalized $E-$functions can be decomposed into the sum of $E-$functions. We distinguish the following three cases.

\subsection{$\Xi^{e\pm}\cdot\Xi^{e\pm}$}\

The product of two $\Xi^{e+}-$ or two $\Xi^{e-}-$functions decomposes into the sum of $\Xi^{e+}$ functions, the signs of the summands are all positive in the first case.  We have the following general formulas of these decompositions which hold for any $\la,\la'\in P$ and $x\in \R^2$
\begin{equation}\label{dec++}
\Xi^{e+}_\la(x)\cdot \Xi^{e+}_{\la'}(x)=\sum_{w\in W^e} \Xi^{e+}_{\la+w\la'}(x),\q \Xi^{e-}_\la(x)\cdot \Xi^{e-}_{\la'}(x)=\sum_{w\in W^e}\sigma^s(w) \Xi^{e+}_{\la+w\la'}(x). 
\end{equation}
The mixed product of $\Xi^{e+}-$ and $\Xi^{e-}-$functions decomposes into the sum of $\Xi^{e-}-$functions:
\begin{equation}\label{dec+-}
\Xi^{e+}_\la(x)\cdot \Xi^{e-}_{\la'}(x)=\sum_{w\in W^e} \sigma^s(w)\Xi^{e-}_{\la+w\la'}(x).
\end{equation}

\begin{example}
Using the explicit form of the even orbits of $C_2$ in Figure \ref{orbitsC2}, we demonstrate the decompositions (\ref{dec++}), (\ref{dec+-}). In the following formulas we choose $\la=(5,3),\,\la'=(1,1)$, omit the variables $(x,y)$ of the $\Xi^{e+}-$ and $\Xi^{e-}-$functions of $C_2$ and write explicitly the products:
\begin{align*}
\Xi^{e+}_{(5,3)}\cdot\Xi^{e+}_{(1,1)}&=\Xi^{e+}_{(6,4)}+\Xi^{e+}_{(2,5)}+ \Xi^{e+}_{(8,1)}+\Xi^{e+}_{(4,2)}\\
\Xi^{e-}_{(5,3)}\cdot\Xi^{e-}_{(1,1)}&=\Xi^{e+}_{(6,4)}-\Xi^{e+}_{(2,5)}- \Xi^{e+}_{(8,1)}+\Xi^{e+}_{(4,2)}\\
\Xi^{e+}_{(5,3)}\cdot\Xi^{e-}_{(1,1)}&=\Xi^{e-}_{(6,4)}-\Xi^{e-}_{(2,5)}- \Xi^{e-}_{(8,1)}+\Xi^{e-}_{(4,2)}.
\end{align*}
\end{example}

\subsection{$\Xi^{s\pm}\cdot\Xi^{s\pm}$}\

The product of two $\Xi^{s+}-$ or two $\Xi^{s-}-$functions decomposes into the sum of $\Xi^{s+}$ functions, the signs of the summands are all positive in the first case.  We have the following general formulas of these decompositions which hold for any $\la,\la'\in P$ and $x\in \R^2$
\begin{equation}\label{decs++}
\Xi^{s+}_\la(x)\cdot \Xi^{s+}_{\la'}(x)=\sum_{w\in W^s} \Xi^{s+}_{\la+w\la'}(x),\q \Xi^{s-}_\la(x)\cdot \Xi^{s-}_{\la'}(x)=\sum_{w\in W^s}\sigma^l(w) \Xi^{s+}_{\la+w\la'}(x). 
\end{equation}
The mixed product of $\Xi^{s+}-$ and $\Xi^{s-}-$functions decomposes into the sum of $\Xi^{s-}-$functions:
\begin{equation}\label{decs+-}
\Xi^{s+}_\la(x)\cdot \Xi^{s-}_{\la'}(x)=\sum_{w\in W^s} \sigma^l(w)\Xi^{s-}_{\la+w\la'}(x).
\end{equation}

\begin{example}
Using the explicit form of the even orbits of $C_2$ in Figure \ref{orbitsC2}, we demonstrate the decompositions (\ref{decs++}), (\ref{decs+-}). In the following formulas we choose $\la=(5,3),\,\la'=(1,1)$, omit the variables $(x,y)$ of the $\Xi^{s+}-$ and $\Xi^{s-}-$functions of $C_2$ and write explicitly the products:
\begin{align*}
\Xi^{s+}_{(5,3)}\cdot\Xi^{s+}_{(1,1)}&=\Xi^{s+}_{(6,4)}+\Xi^{s+}_{(2,4)}+ \Xi^{s+}_{(8,2)}+\Xi^{s+}_{(4,2)}\\
\Xi^{s-}_{(5,3)}\cdot\Xi^{s-}_{(1,1)}&=\Xi^{s+}_{(6,4)}-\Xi^{s+}_{(2,4)}- \Xi^{s+}_{(8,2)}+\Xi^{s+}_{(4,2)}\\
\Xi^{s+}_{(5,3)}\cdot\Xi^{s-}_{(1,1)}&=\Xi^{s-}_{(6,4)}-\Xi^{s-}_{(2,4)}- \Xi^{s-}_{(8,2)}+\Xi^{s-}_{(4,2)}.
\end{align*}
\end{example}

\subsection{$\Xi^{l\pm}\cdot\Xi^{l\pm}$}\

The product of two $\Xi^{l+}-$ or two $\Xi^{l-}-$functions decomposes into the sum of $\Xi^{l+}$ functions, the signs of the summands are all positive in the first case.  We have the following general formulas of these decompositions which hold for any $\la,\la'\in P$ and $x\in \R^2$
\begin{equation}\label{decl++}
\Xi^{l+}_\la(x)\cdot \Xi^{l+}_{\la'}(x)=\sum_{w\in W^l} \Xi^{l+}_{\la+w\la'}(x),\q \Xi^{l-}_\la(x)\cdot \Xi^{l-}_{\la'}(x)=\sum_{w\in W^l}\sigma^s(w) \Xi^{l+}_{\la+w\la'}(x). 
\end{equation}
The mixed product of $\Xi^{l+}-$ and $\Xi^{l-}-$functions decomposes into the sum of $\Xi^{l-}-$functions:
\begin{equation}\label{decl+-}
\Xi^{l+}_\la(x)\cdot \Xi^{l-}_{\la'}(x)=\sum_{w\in W^l} \sigma^s(w)\Xi^{l-}_{\la+w\la'}(x).
\end{equation}
\begin{example}
Using the explicit form of the even orbits of $C_2$ in Figure \ref{orbitsC2}, we demonstrate the decompositions (\ref{decl++}), (\ref{decl+-}). In the following formulas we choose $\la=(5,3),\,\la'=(1,1)$, omit the variables $(x,y)$ of the $\Xi^{l+}-$ and $\Xi^{l-}-$functions of $C_2$ and write explicitly the products:
\begin{align*}
\Xi^{l+}_{(5,3)}\cdot\Xi^{l+}_{(1,1)}&=\Xi^{l+}_{(6,4)}+\Xi^{l+}_{(6,1)}+ \Xi^{l+}_{(4,5)}+\Xi^{l+}_{(4,2)}\\
\Xi^{l-}_{(5,3)}\cdot\Xi^{l-}_{(1,1)}&=\Xi^{l+}_{(6,4)}-\Xi^{l+}_{(2,4)}- \Xi^{l+}_{(4,5)}+\Xi^{l+}_{(4,2)}\\
\Xi^{l+}_{(5,3)}\cdot\Xi^{l-}_{(1,1)}&=\Xi^{l-}_{(6,4)}-\Xi^{l-}_{(6,1)}- \Xi^{l-}_{(4,5)}+\Xi^{l-}_{(4,2)}.
\end{align*}
\end{example}

\section{Concluding Remarks}
\begin{itemize}
 \item 
The short roots $W\Delta^s$ of the Weyl group of $C_2$ ($G_2$) is of the type $A_1\times A_1$ ($A_2$). This leads to the fact, that in the case of $C_2$ the functions $\Xi^{s+}$ can be identified with $C$-orbit functions of the group of $A_1\times A_1$ and $\Xi^{s-}$ with the $S$-orbit functions of $A_1\times A_1$. In particular, the function $\Xi^{s+}_{(a,b)}$ of $C_2$ differs only by rotation by $\pi/4$ from the $C$-function $\varphi_{a+b,b}$ of $A_1\times A_1$ (and similarly for the $\Xi^{s-}$). In the case of $G_2$, the functions $\Xi^{s+}_{(a,b)}$ differ from the functions $\varphi_{(a,a+b)}$ of $A_2$ by the rotation by $\pi/3$ (and similarly for the $\Xi^{s-}$).

\medskip

\item
For all groups \eqref{thegroups} there are precisely six families of $E$-function analogous to the rank 2 cases here.  The kernels of defining homomorphisms are the symmetry groups of the subsystems of roots of the same length. Namely,
\begin{gather}\label{subroots}
W\Delta^s\text{ is of type }\begin{cases}
  &nA_1\ \text{in}\ B_n\\
  &D_n\  \text{in}\ C_n\\
  &D_4\  \text{in}\ F_4\\  \end{cases}\,,\qquad\qquad
W\Delta^l\text{ is of type }\begin{cases}
  &D_n\  \text{in}\ B_n\\
  &nA_1\ \text{in}\ C_n\\
  &D_4\  \text{in}\ F_4\\  \end{cases}\,,
\end{gather}
where $nA_1$ denotes the semisimple Lie algebra, with $A_1\times\cdots\times A_1$ multiplied $n$ times. In \eqref{subroots} we use the known isomorphisms $D_2=A_1\times A_1$,  $D_3=A_3$, and  $B_2=C_2$.
\medskip
\item
The most frequently analyzed 2-dimensional digital data are sampled on rectangular domains. Such data need first to be placed in $F_M$ with appropriate choice of the integer $M$ to match the density of the data lattice and the lattice points in $F_M$. That is done more efficiently when $F_M$ matches more closely the shape of the data lattice. The functions on Figures 4, 8, and 10 illustrate the new choices one did not have so far.

\medskip

\item
For now unexplored remain the properties of the $E$-functions of all the types when the dominant weight of the function is a point in $\R^n$ but not a point of the weight lattice $P\subset\R^n$. Definitions and many properties of such functions are analogs of the properties of the $E$-functions described here. Obvious application of such functions are Fourier integral expansions as opposed to Fourier series.

\medskip

\item
Weyl group orbit functions are closely related to multivariable orthogonal polynomials. That relation exists also in the case of $E$-functions of all the types. In our opinion this relation merits an explicit description.

\medskip

\item
Very little is known about arithmetic properties of the $E$-functions, except of those facts that can be readily deduced from the rich arithmetic properties of the characters \cite{MP84},\cite{MPS83}. 
\medskip
\end{itemize}

\section*{Acknowledgments}
We gratefully acknowledge the support of this work by the Natural Sciences and Engineering Research Council of Canada and by the Doppler Institute of the Czech Technical University in Prague. JH is grateful for the hospitality extended to him at the Centre de recherches math\'ematiques, Universit\'e de Montr\'eal. JH gratefully acknowledges support by the Ministry of Education of Czech Republic (project MSM6840770039). JP expresses his gratitude for the hospitality of the Doppler Institute.

\end{document}